\documentclass[11pt]{article}

\usepackage[T1]{fontenc}
\usepackage[english]{babel}

\usepackage{amsmath,amssymb,amsthm}

\usepackage[letterpaper]{geometry}
\usepackage{float}
\usepackage{setspace}
\usepackage{tikz}
\usetikzlibrary{backgrounds}

\usepackage[authoryear]{natbib}
\usepackage[pdfusetitle]{hyperref}
\hypersetup{
  colorlinks=true,
  linkcolor=blue,
  citecolor=blue,
  urlcolor=blue
}

\theoremstyle{definition}
\newtheorem{defn}{Definition}
\newtheorem{example}{Example}
\newtheorem*{example*}{Example}

\theoremstyle{plain}
\newtheorem{thm}{Theorem}
\newtheorem{cor}{Corollary}
\newtheorem{prop}{Proposition}
\newtheorem{lem}{Lemma}

\theoremstyle{remark}
\newtheorem{claim}{Claim}
\newtheorem{rem}{Remark}
\newtheorem*{rem*}{Remark}

\title{Revealed Preferences of One-Sided Matching}
\author{Andrew Tai\thanks{%
  A majority of this work was done while the author was a PhD student at UC Berkeley. I am indebted
  to David Ahn, Federico Echenique, Haluk Ergin, Shachar Kariv, and Chris Shannon for their guidance.
  I also thank seminar participants at UC Berkeley, WashU EGSC, and Cal Poly SLO for helpful
  comments. All errors are my own.
}}
\date{September 2026}

\begin{document}
\maketitle
\begin{abstract}
  \noindent I find the testable implications of the core in an exchange economy with unit demand
  when agents' preferences are unobserved. To do so, I develop a model of aggregate matchings in
  which the core is falsifiable; the identifying assumption is that agents' preferences are solely
  determined by observable characteristics. I find conditions that characterize when observed
  economies are compatible with the core. These conditions are meaningful, intuitive, and tractable,
  providing a nonparametric test for the core in the style of revealed preferences.
\end{abstract}

\section{Introduction}

This paper studies the testable implications of the core in exchange economies with indivisible
goods and unit demand. The setting coincides with the house-swapping model of \citet{SS74}. As in
classical revealed preference theory, I take agents, endowments, and allocations to be observable,
but preferences to be unobserved. Given such data, I investigate the testable implications of the
core. In both models with and without monetary transfers, I find conditions that characterize when
the observables are consistent with the core (``rationalizability''). That is, these conditions can
falsify the market being in the equilibrium.

The exchange economy is a foundational model in economics for situations without explicit
production. With unit demand and indivisible goods, these models correspond to exchanges or
allocations of ``large'' objects. \citeauthor{SS74} refer to these large indivisible goods as
``houses''; indeed, this is interpretable as a stylized model of housing allocation. The model is
often applied to settings such as living donor organ exchange, school assignment, and course
allocation. The process of allocation may be unknown or ambiguous. We may not observe any of the
series of trades, and even in the setting with monetary transfers, competitive prices are not
inherent to the model. The allocation process can be decentralized trade, as in a Walrasian market;
or via a centralized mechanism.

An example of a market with many of these attributes is the Singaporean public housing
market.\footnote{\href{https://web.archive.org/web/20220119024114/https://www.singstat.gov.sg/-/media/files/publications/population/population2021.pdf}{Population
Trends, Department of Statistics, Ministry of Trade \& Industry, Republic of Singapore} -- This is
also an important market; 79\% of Singaporeans live in public housing.} The government allocates new
public housing quarterly via a centralized build-to-order mechanism. Applicant households submit
interest in a new development and are awarded a subsidized 99 year lease, which they have the right
to sell after five years. This setting incorporates many of the features described above; housing is
a large good, initial prices are restricted, and owners have trading rights afterward. It is also
plausible that households with the same observable characteristics share the same preferences over
developments.

The core is a natural equilibrium notion for this setting. Informally, it captures group stability
by requiring that no coalition would prefer to break off and re-trade their endowments among
themselves. Alternatively, any beneficial trades have already been made. Implicitly, these
coalitions can plausibly find each other; thus it is a sensible equilibrium notion for a market that
is ``small'' relative to the ``large'' goods. The core is also Pareto efficient. Importantly, the
core does not require prices, which are not inherent to this model. However, I also present
equivalence results for the core and competitive equilibrium in this model.

The conditions I present characterize restrictions on the observable data of core allocations. An
analyst may wish to check for the core for a few reasons. Equilibrium itself may be the object of
interest -- an economy which satisfies the conditions is plausibly stable and Pareto optimal. Other
analysis may also require the market to be in equilibrium, such as study of the preferences. The
conditions for rationalizability also provide ex ante predictions for equilibrium market outcomes.
More philosophically, this paper addresses the scientific content of the core. That is, I develop
reasonable conditions under which the core is falsifiable, akin to the falsifiability of utility
maximization in classical revealed preferences.

There is at least one other way to interpret my results. Observers may deal with settings where the
centralized mechanism is unknown and therefore cannot be directly evaluated. In practice, many
mechanisms are hidden, or no particular mechanism is used at all, such as administrators exercising
personal judgment to determine allocations. But we nevertheless want to determine whether these
unknown mechanisms might be stable. \citet{Grigoryan} develop a theory of \emph{auditability}, where
mechanism implementers may deviate for various reasons; auditability measures how much information
the participants need to detect a deviation. This paper offers a way to evaluate mechanisms when
essentially nothing is known about the matching process, but the analyst still wants to determine
whether the allocation is may be stable. Alternatively, there may be no centralized mechanism at
all. In this interpretation, I develop a theory to test stability when there is no particular
matching process.

To rationalize a market, it is sufficient to find a preference profile such that it is in the core.
However, this is an existential condition, requiring a combinatorial number of cases. I instead find
feasible conditions inspired by revealed preference theory.

In classical consumer demand revealed preference theory, we infer that the chosen option is the best
among affordable options. \citet{Afriat67} then proceeds from here to construct utility values.
However, in an exchange economy, the available options are not exogenously determined by some
budget. Stability in an exchange market is determined by all other agents' preferences. Further, the
core is not equivalent to maximizing social utility, even when we allow monetary transfers.

To gain traction in this setting, I deal with aggregate matchings, akin to \citet{CS06}'s empirical
work in marriage markets. Objects are grouped into types, equivalent within type and distinct across
types. For instance, these may be apartments in the same development or houses in the same
neighborhood, which can be regarded as essentially the same. I also assume that agents can be binned
into ``types'' with the same preference, analogous to the assumption of \citet{ELSY13}. Stated
another way, agents with the same observable characteristics (such as age, wealth, and
socioeconomics) have the same preferences. This is a strong assumption as it rules out individual
heterogeneity.\footnote{In the model without transfers, rankings are purely ordinal, so small
cardinal heterogeneity is allowed.} However, allowing for enough individual heterogeneity also
allows any observed market to be rationalized.\footnote{Simply declare all agents' allocations to be
their favorite things.} In exchange, the resulting test for rationalizability is nonparametric, in
the spirit of revealed preferences.

I introduce a graph representation of exchange economies and develop graph theoretic results around
it. This construction is extremely tractable, and it gives rise to intuitively appealing conditions
for rationalizability. The graph construction's tractability also suggests ways to develop
``smoother'' definitions and statistical tests of rationalizability, which I discuss briefly in the
conclusion. In the setting without transfers, rationalizability is equivalent to equal treatment
within each type in each market segment. In the setting with monetary transfers, there are two
equivalent conditions: the existence of a price vector rationalizing the allocation as a competitive
equilibrium; and a cyclic monotonicity condition similar to many in the revealed preferences
literature.

This paper contributes to the study of the testable implications of equilibria. The
Sonnenschein--Mantel--Debreu theorem \citeyearpar{Sonnenschein72,Mantel74,Debreu74} gives a famous
``anything goes'' result on the excess demand function in competitive equilibrium. In the same vein,
\citet{MasColell77} shows that there are essentially no restrictions on rationalizable prices in
competitive equilibria. \citet{BrownMatzkin96} apply revealed preference theory to obtain
restrictions on competitive equilibrium outcomes when a series of markets is observed.
\citet{BossertSprumont02} find conditions for core rationalizability in a two agent economy with
divisible commodities. I study a distinct setting -- exchange economies with indivisible goods and
unit demand -- and find tractable and intuitive restrictions on core allocations.

Additionally, I contribute to the growing literature on the revealed preferences of matching.
\citeauthor{ELSY13} study the revealed preferences of matching in marriage markets with aggregate
matching type data. \citet{Echenique08} finds testable implications for two-sided matching when
individuals participate in a series of markets; \citet{DS22} find equivalent conditions.

\section{Model}

I first present the model and notation for the case without monetary transfers. The following
section presents the additions for the case of monetary transfers.

\subsection{Without transfers}\label{subsec:modelNT}

The basis of the model is the \citeauthor{SS74} house-swapping model with the addition that objects
and agents are grouped into types. Agents of the same type share the same (unobserved to the
analyst) preference. Let the set of agent types as $A=\left\{ 1,2,...,A\right\} $, where $A$ denotes
both the set and its cardinality at minimal risk of confusion; let $A<\infty.$ Denote the set of
individual agents as $\mathcal{A}=\left\{ 1a,1b,...;2a,2b,...;Aa,Ab,...\right\} $, and let
$\left|\mathcal{A}\right|<\infty$. Notationally, $\mathcal{A}$ also encodes the types of each
individual; e.g., $1a$ and $1b$ are two individuals of the same type $1$. I will refer to $i\in A$
as a ``type'', and $ik\in\mathcal{A}$ as an ``individual'' or ``agent''.

Denote the set of object types $H$, also with cardinality $H$. I denote each object type as a unit
vector in $\mathbb{R}^{H}$; that is,
\[
  H=\{\underset{:=h_{1}}{\underbrace{\left(1,0,...,0\right)}},\underset{:=h_{2}}{\underbrace{\left(0,1,0,...,0\right)}},\underset{:=h_{H}}{\underbrace{\left(0,...,0,1\right)}}\}\subset\mathbb{R}^{H}
\]
It will not be necessary to refer to individual objects -- i.e., there is no object analogue of
$\mathcal{A}$.

Each agent is endowed with an object, denoted $e_{ik}\in H$. An endowment vector is
$e=\left(e_{ik}\right)_{ik\in\mathcal{A}}$. An allocation is
$x=\left(x_{ik}\right)_{ik\in\mathcal{A}}$ such that
$\sum_{ik\in\mathcal{A}}x_{ik}=\sum_{ik\in\mathcal{A}}e_{ik}$. That is, the number of allocated
objects of each type is equal to the number supplied.

A feasible sub-allocation for a coalition $A'\subseteq\mathcal{A}$ is
$x'=\left(x_{ik}'\right)_{ik\in A'}$ such that $\sum_{ik\in A'}x_{ik}'=\sum_{ik\in A'}e_{ik}$.

Each type $i$ has a \emph{strict} preference $\succsim_{i}$ over $H$; all $ik$ of type $i$ have the
same preference. I will discuss this more in Section \ref{subsec:Model-Discuss}. Denote
$\succsim=\left(\succsim_{i}\right)_{i\in A}$ be the preference profile. With minimal risk (or
consequence) of confusion, this could also be the profile of agents
$\succsim=\left(\succsim_{ik}\right)_{ik\in\mathcal{A}}$.

The equilibrium concept used in this paper is the strict core.
\begin{defn}
  A \textbf{weak blocking coalition} is $A'\subseteq\mathcal{A}$ with feasible sub-allocation $x'$
  such that $x_{ik}'\succsim_{i}x_{ik}$ for all $ik\in A'$, and $x_{ik}'\succ_{i}x_{ik}$ for at
  least one $ik\in A'$. An allocation $x$ is in the \textbf{strict} \textbf{core} for a preference
  profile $\succsim$ if there is no weak blocking coalition.
\end{defn}
By convention, when a blocking coalition $A'$ is one individual, I say $x$ is not individually
rational.\footnote{A blocking coalition of one individual $ik$ means $e_{ik}\succ_{i}x_{ik}$.}

We can now state the main objective of the paper. If we observe individuals, types, endowments, and
allocations, could the market be in the core? Formally, is there a preference profile such that $x$
is in the strict core?
\begin{defn}
  A tuple $\left(A,\mathcal{A},H,e,x\right)$ is an \textbf{NT-economy} (non-transfers-economy). An
  economy is \textbf{NT-rationalizable} if there exists a preference profile $\succsim$ such that
  $x$ is in the strict core.
\end{defn}

\subsection{With transfers}

I now introduce monetary transfers. The notation for types, agents, and objects remains the same.
Endowments are now an object and amount of money,
$(e,\omega)=\left(e_{ik},\omega_{ik}\right)_{ik\in\mathcal{A}}$, where $e_{ik}\in H$ and
$\omega_{ik}\in\mathbb{R}_{++}$. Likewise, an allocation is an object and amount of money
$(x,m)=\left(x_{ik},m_{ik}\right)_{ik\in\mathcal{A}}$, such that $m_{ik}\in\mathbb{R}_{++}$,
$\sum_{ik\in\mathcal{A}}x_{ik}=\sum_{ik\in\mathcal{A}}e_{ik}$, and
$\sum_{ik\in\mathcal{A}}m_{ik}\leq\sum_{ik\in\mathcal{A}}\omega_{ik}$. Note that endowed and
allocated money are restricted to be strictly positive. Analogously, a feasible sub-allocation for a
coalition $A'$ is $(x',m')=\left(x_{ik}',m_{ik}'\right)_{ik\in A'}$ such that
$\sum_{ik\in A'}x_{ik}'=\sum_{ik\in A'}e_{ik}$ and
$\sum_{ik\in A'}m_{ik}'\leq\sum_{ik\in A'}\omega_{ik}$.

Let utility $V_{i}:H\times\mathbb{R}_{+}\rightarrow\mathbb{R}$ be quasilinear, given by
$V_{i}(h,m)=v_{i}(h)+m$. Notice that the subscript is on types. The $v_{i}(\cdot)$ can be
interpreted as a utility index over $H$; that is, it is an $H$-dimensional vector of real numbers
representing cardinal utility for objects. We can regard this model as a partial equilibrium
analysis, where all other goods are grouped into money. This is also a common assumption in market
design and matching (e.g. \citealp{GulPesendorferZhang}).

The equilibrium concept in the transfers model is the weak core.
\begin{defn}
  For an allocation $(x,m)$, a strong blocking coalition is $A'\subseteq\mathcal{A}$ with feasible
  sub-allocation $(x',m')|_{A'}$ such that $V_{i}(x'_{ik},m'_{ik})>V_{i}(x_{ik},m_{ik})$ for all
  $ik\in A'$. An allocation $(x,m)$ is in the \textbf{weak core} for $\left(v_{i}\right)$ if there
  is no strong blocking coalition.
\end{defn}
The weak core and strict core coincide in most cases, as any strictly better off members can give
$\varepsilon$ payments to any indifferent members. The exception is when all strictly better off
members exhaust their money in a candidate blocking coalition. The assumption that
$\omega_{ik},m_{ik}>0$ ensures that money truly enters the model and that the weak core and strict
core coincide for rationalizable allocations.\footnote{It can be argued as in \citet{Kaneko82} and
\citet{Quinzii84} that money is a bundle of goods outside the model, and it is not ``normal'' to
consume only one indivisible good.}

The definition of rationalizability is analogous. The analyst observes individuals, types,
endowments, and allocations (the latter two including money). I seek a preference profile such that
$(x,m)$ is in the core.
\begin{defn}
  A tuple $\left(A,\mathcal{A},H,(e,\omega),(x,m)\right)$ is a \textbf{T-economy}
  (transfers-economy). An economy is \textbf{T-rationalizable} (transfers-rationalizable) if there
  exists utility indexes $(v_{i})$ such that $(x,m)$ is in the weak core. It is \textbf{strictly
  T-rationalizable} if it is T-rationalizable with some strict utility indexes; that is,
  $v_{i}(h)=v_{i}(h')$ if and only if $h=h'$ for all $i$.
\end{defn}
The main result for T-economies will deal with T-rationalizability; I do not impose that the utility
indexes $\left(v_{i}\right)$ are strict over $H$. I will discuss afterwards how strict
T-rationalizability is a corollary of the main result.

\begin{table}
  \centering
  \caption{Notation}
  \label{tab:Notation}

  \begin{tabular}{cccc}
    Object             & Without transfers & With transfers          & Generic member \\
    \hline
    Agent types        & $A$               & $A$                     & $i$            \\
    Individuals/agents & $\mathcal{A}$     & $\mathcal{A}$           & $ik$           \\
    Objects            & $H$               & $H$                     & $h$            \\
    Endowment          & $e$               & $(e,\omega)$            &                \\
    Allocation         & $x$               & $(x,m)$                 &                \\
    Preferences        & $\succsim$        & $V_{i}(h,m)=v_{i}(h)+m$ &                \\
  \end{tabular}
\end{table}

\subsection{Discussion}\label{subsec:Model-Discuss}

This paper derives necessary and sufficient conditions for an economy to be rationalizable. That is,
I characterize allocations which are compatible with the core. As mentioned earlier, this
characterization can be used to check for equilibrium; this may be of interest in and of itself or
be necessary for further analysis.

Mechanically, this question is the reverse direction of the classic house-swapping question. We have
a house-swapping market as in \citet{SS74} where there are potentially multiple copies of each
object. Given an allocation, we are seeking preferences generating it.

The key identifying assumption is common preferences within agent type. This introduces discipline
to the problem. As noted above, this gives the economy testable content; with enough individual
heterogeneity, any economy is rationalizable.\footnote{There are alternatives, such as repeated
re-matchings as in \citet{Echenique08}.} While not explicitly modeled, this is akin to an assumption
that preferences are solely functions of agents' observable characteristics. If there are observable
traits of agents $X_{a}$ and of objects $X_{h}$, rankings are generated by some utility function
$u(X_{a},X_{h})$. For the non-transfers case, the resulting characterizations are completely
nonparametric. For transfers case, I impose quasilinear utility; but the utility for objects
$v_{i}(h)$ is otherwise nonparametric. Since the non-transfers preferences are purely ordinal, some
\emph{cardinal} heterogeneity is allowed, as long as the same \emph{ordinal} rankings are generated.

If types are constructed from binned variables, the analyst has some degree of choice. Coarser bins
result in stronger implications on the allocation, and finer bins result in weaker implications. The
``correct'' tradeoff is outside of the model of this paper, but the analyst can decide on the most
reasonable choice.

Finally, rationalizability is a meaningful concept; it is not hard to construct economies that are
not rationalizable. For example, two agents with the same preferences should never trade their
endowments.

\subsection{Graphs}

I represent economies in graph-theoretic terms in order to take advantage of results from graph
theory and to parsimoniously present the main results. I first introduce some standard definitions
for directed graphs that will be useful. Familiar readers can skip this subsection.
\begin{defn}
  A \textbf{directed graph (digraph)} is $D=(V,E)$, where $V$ is the set of vertices, and $E$ is the
  set of arcs. An \textbf{arc} is an sequence of two vertices $(v_{i},v_{j})$; here I allow for arcs
  of the form $(v_{i},v_{i})$, called a self-loop.\footnote{This is more formally called a
  \textbf{directed pseudograph}.} A \textbf{$(v_{1},v_{k})$-path} is sequence of vertices
  $(v_{1},v_{2},...,v_{k})$ where each $v_{i}$ is distinct, and $(v_{i-1},v_{i})\in E$ for each
  $i\in\{2,...,k\}$. A \textbf{cycle} is a sequence of vertices $(v_{1},v_{2},...,v_{k},v_{1})$,
  where each $v_{i}$ is distinct except for the first and last, and $(v_{i-1},v_{i})\in E$ for each
  $i\in\{2,...,k\}$. I will also include self-loops $(v_{1},v_{1})$ as cycles. Equivalently, a path
  is a sequence of arcs $\left((v_{1},v_{2}),...,(v_{k-1},v_{k})\right)$, and analogously for
  cycles. The \textbf{indegree} of a vertex $d^{-}(v_{i})=\left|v_{j}:(v_{j},v_{i})\in E\right|$ is
  the number of arcs pointing at $v_{i}$. Likewise, the \textbf{outdegree} of a vertex
  $d^{+}(v_{i})=\left|v_{j}:(v_{i},v_{j})\in E\right|$ is the number of arcs pointing from $v_{i}$.
\end{defn}
\begin{defn}
  A \textbf{weighted directed graph} is $D=\left(V,E,\ell(\cdot)\right)$, where $(V,E)$ is a
  directed graph, and $\ell:E\rightarrow\mathbb{R}$ is the \textbf{length} (or \textbf{weight})
  function over arcs. The length of a path or cycle $(v_{1},v_{2},...,v_{k})$ is
  $\sum_{i=1}^{k-1}\ell(v_{i},v_{i+1})$.
\end{defn}
The next definition is used in the main result and its discussion.
\begin{defn}
  A \textbf{strongly connected component (SCC)} of a digraph $D=(V,E)$ is a maximal set of vertices
  $S\subseteq V$ such that for all distinct vertices $v_{i},v_{j}\in S$, there is a
  $(v_{i},v_{j})$-path and a $(v_{j},v_{i})$-path. By convention, there is always a path from
  $v_{i}$ to itself, even if $(v_{i},v_{i})\not\in E$; an isolated vertex is an SCC.
\end{defn}
Informally, an SCC is a maximal set of vertices such that there is a path from any vertex to any
vertex.

\begin{figure}[H]
  \centering
  \caption{Example of strongly connected components}
  \label{fig:scc-ex}


\begin{tikzpicture}[
roundnode/.style={circle, draw=black, fill=white, very thick, minimum size=4mm}, 
scale=0.7
]

\node[roundnode]    (2)    at (0,3)   {} ;
\node[roundnode]    (1)    at (0,0)   {};
\node[roundnode]    (3)    at (-1.5,1.5)  {} ;

\node[roundnode]    (1a)    at (3,3)  {} ;
\node[roundnode]    (4)    at (3,0)  {}  ;

\node[roundnode]    (1b)    at (6,3)   {} ;

\path [thick, ->, shorten >=3pt, shorten <=3pt]
    (1) edge [bend right] (2) 	
    (2) edge [bend right] (3) 
    (3) edge [bend right] (1) 
    (1a) edge [bend right] (4) 	
    (4) edge [bend right] (1a) 
    (1a) edge [bend right] (2) 	
    (1a) edge [bend left] (1b) ;	

\begin{pgfonlayer}{background}
	\filldraw [fill=black!20] 	
	(-2.5, -1) rectangle (1, 4);
	
	\filldraw [fill=black!20] 
	(2, -1) rectangle (4, 4);
	
	\filldraw [fill=black!20] 
	(5, 2) rectangle (7, 4);

\end{pgfonlayer}

\end{tikzpicture}
\end{figure}
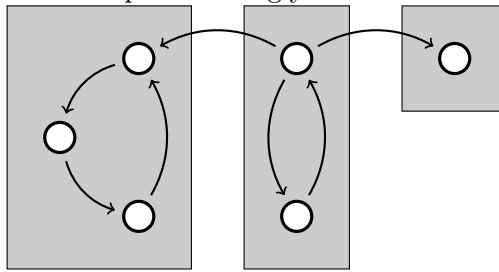

The vertices of any digraph can be uniquely partitioned into SCCs. An algorithm by \citet{Tarjan}
finds a partition in linear time, $O\left(|V|+|E|\right)$. Figure \ref{fig:scc-ex} illustrates such
a partition; the SCCs are shaded. For example, in the left-most SCC, there is a path from any vertex
to any other vertex. It is also maximal, since including other vertices destroys this property.

\section{Rationalizability}

In this section, I present necessary and sufficient conditions for an economy to be rationalizable.
I will first present the graph representation of economies, which I use to show the result for
NT-economies. I will then present the analogous results for T-economies.

\subsection{Without transfers}

First, I introduce a graph construction that is important for the main results. Construct the
digraph $\mathcal{G}_{NT}\left(A,\mathcal{A},H,e,x\right)=(\mathcal{A},E)$ as follows: each
individual is a vertex. Draw arcs from $ik$ to \emph{all} vertices $i'k'$ that are endowed with
$x_{ik}$. That is, let $(ik,i'k')\in E$ if $x_{ik}=e_{i'k'}$. The following example illustrates.
\begin{example}
  \label{exa:NTexample}
  Consider the economy described below.

  \[
    \begin{array}{ccc}
      ik & e_{ik} & x_{ik} \\
      1a & h_{1}  & h_{2}  \\
      1b & h_{2}  & h_{2}  \\
      1c & h_{4}  & h_{5}  \\
      2a & h_{2}  & h_{3}  \\
      2b & h_{5}  & h_{4}  \\
      3a & h_{3}  & h_{1}
    \end{array}
  \]
  That is, $e_{1b}=e_{2a}$, and other endowments are unique. The graph $\mathcal{G}_{NT}$ is given
  below in Figure \ref{fig:ExampleNT-fig}.

  \begin{figure}[h]
    \centering
    \caption{Figure for Example \ref{exa:NTexample}}
    \label{fig:ExampleNT-fig}


\begin{tikzpicture}[
roundnode/.style={circle, draw=black, fill=black!5, thick, minimum size=5mm}, 
scale=0.7
]

\tikzstyle{every node}=[font=\scriptsize]

\node[roundnode]    (2a)    at (0,3) 	 	{$2a/h_2$};
\node[roundnode]    (1a)    at (0,0)   		{$1a/h_1$};
\node[roundnode]    (3a)    at (-1.5,1.5) 	{$3a/h_3$};

\node[roundnode]    (1b)    at (3,3)		{$1b/h_2$};

\node[roundnode]    (1c)    at (6,3)		{$1c/h_4$};
\node[roundnode]    (2b)    at (6,0)		{$2b/h_5$};

\path [thick, ->, shorten >=3pt, shorten <=3pt]
    (1a) edge [bend right] (2a) 	
    (2a) edge [bend right] (3a) 
    (3a) edge [bend right] (1a) 
    (1b) edge [loop right] (1b) 	
    (1b) edge [bend right] (2a) 	
    (1a) edge [bend right] (1b)
    (1c) edge [bend right] (2b) 	
    (2b) edge [bend right] (1c) ;

\end{tikzpicture}
  \end{figure}
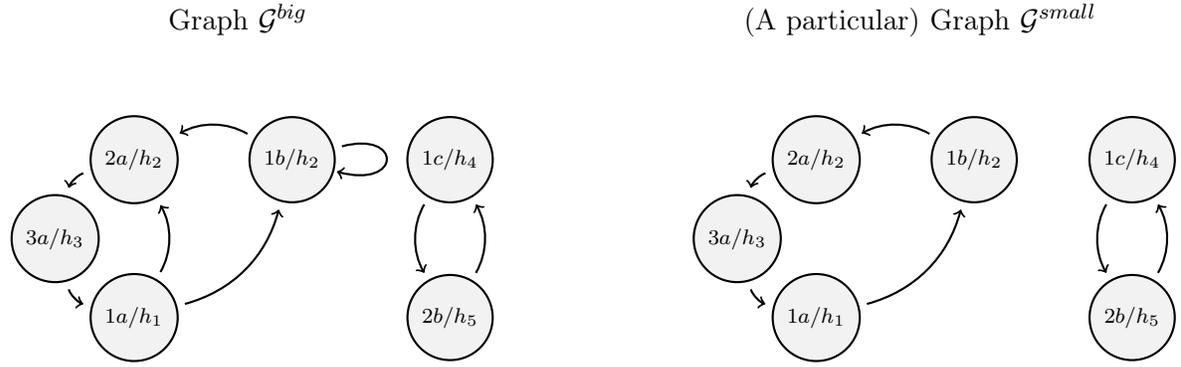
\end{example}
The SCCs of $\mathcal{G}_{NT}$ are the focus of the main result. In the context of this paper's
setting, the SCCs are interpretable as partitioning the market into segments that trade among
themselves. I will refer to these informally as \emph{market segments}. Readers familiar with
matching may recognize that any allocation can be partitioned into trading cycles.\footnote{Not
necessarily Gale's TTC cycles -- no claim on optimality is made here yet.} In the setting of
\citeauthor{SS74} without indifferences, this partition is unique. In the present setting, these
cycles may not be unique; however, $\mathcal{G}_{NT}$ superimposes all such trading cycles onto one
graph.

I now state the main result for the NT-economy.
\begin{thm}
  \label{thm:Main}
  Fix an NT-economy $\left(A,\mathcal{A},H,e,x\right)$, and consider
  $\mathcal{G}_{NT}\left(A,\mathcal{A},H,e,x\right)$. The economy is NT-rationalizable if and only
  if: for agents of the same type $ik,ik'$ in the same SCC $S$, $x_{ik}=x_{ik'}$. That is, if
  $ik,ik'\in S$ are the same type and in the same SCC, they receive the same object type.
\end{thm}
\begin{proof}
  Appendix.
\end{proof}
The full proof is contained in the appendix. I give a sketch of the proof below.
\begin{proof}[Proof sketch of Theorem \ref{thm:Main}]
  A key feature of $\mathcal{G}_{NT}$ is that all objects of the same type are contained in the same
  SCC. The proof of this claim is under Corollary \ref{cor:SCCalltypes}.

  To prove ``if'': First, find the decomposition of $\mathcal{G}_{NT}$ into SCCs. Then assign an
  arbitrary order to the SCCs, and assign preferences in this order. In the first SCC $S_{1}$, set
  all types' top rank to be the objects they receive. By assumption, all agents of the same type in
  the same SCC receive the same object, so this is a well defined procedure. In the second SCC
  $S_{2}$, there may be types who were not present in $S_{1}$; let these types' top rank be the
  objects they receive. For types who were present in $S_{1}$, set their second ranked object to be
  what they receive. Then continue through the remaining SCCs in this way. Since all objects of the
  same type are in the same SCC, the procedure never attempts to ``re-assign'' a preference in a
  later step. That is, objects never ``re-appear'' after being assigned to a preference rank the
  first time. The argument that this creates no blocking coalitions is similar to the argument
  behind Gale's proof for TTC.

  To prove ``only if'', I show that when the condition is violated, there is a blocking coalition
  for all preference profiles. One of the two agents of the same type must be worse off; this one
  can form a blocking coalition with a subset of other members of the SCC.
\end{proof}
\begin{example*}[Example \ref{exa:NTexample} continued.]
  The $\mathcal{G}_{NT}$ has two SCCs: the left component and the right component. To apply the
  theorem, select one order arbitrarily. Let the left component be $S_{1}$, and the right be
  $S_{2}$. Let $\succsim_{i}(k)$ denote type $i$'s $k^{th}$ favorite object.
  \begin{enumerate}
    \item In $S_{1}$, assign all agents' $\succsim_{i}(1)=\mu(i)$, so
          \[
            \begin{array}{cc}
              i & \succsim_{i}(1) \\
              1 & h_{2}           \\
              2 & h_{3}           \\
              3 & h_{1}
            \end{array}
          \]
    \item In $S_{2}$, assign all agents' $\succsim_{i}(1)=\mu(i)$ for any $i$ who were not in
      $S_{1}$. (Here, both types 1 and 2 were present in $S_{1}$.) Otherwise, let
      $\succsim_{i}(2)=\mu(i)$.
          \[
            \begin{array}{cc}
              i & \succsim_{i}(2) \\
              1 & h_{5}           \\
              2 & h_{4}
            \end{array}
          \]
    \item Assign remaining preferences arbitrarily (omitted).
  \end{enumerate}
  To check for a blocking coalition, observe that all agents in $S_{1}$ all receive their favorite
  objects. Only agents in $S_{2}$ are unsated. Then in any candidate blocking coalition $(A',\mu')$,
  we require $\mu'(1c)=h_{2}$ or $\mu'(2b)=h_{3}$. This requires at least one agent in
  $A'\cap S_{1}$ to receive either $h_{4}$ or $h_{5}$, which are strictly dispreferred.
\end{example*}
The condition required in Theorem \ref{thm:Main} is easy to check; Tarjan's algorithm finds the
partition into SCCs in linear time. Within each SCC, checking for a non-repeated agent type-object
type pair is linear in the number of agents.

\subsection{Discussion and related results}

The most direct interpretation of Theorem \ref{thm:Main} is the following: whenever agents with the
same preferences are in the same market segment, they receive the same object type. Informally,
agents in the same market segment have similarly desirable endowments; if there are multiple agents
of the same type in an SCC, one should not be worse off. Within a market segment, any agent can make
a series trades to receive any object in this segment; the formal proof of Theorem \ref{thm:Main}
uses this idea this to find a blocking coalition.

A second interpretation is in the context of a competitive equilibrium market.\footnote{I mean the
usual definition. I give the formal definition in Definition \ref{def:CE} in the appendix.}
\citet{RP77} show that any strict core allocation is a competitive equilibrium allocation in the
typical object exchange setting with no indifferences. \citet{Wako84} establishes that a strict core
allocation is also a competitive equilibrium allocation in the setting with indifferences. If $x$ is
in the core for some preference profile $\succsim$, it is also a competitive equilibrium allocation
for some price vector. A supporting price vector is descending in the (arbitrarily selected) order
of SCCs. Thus if two agents are in the same SCC, their endowments are worth the same in competitive
equilibrium. The necessity of the condition becomes immediate; two agents with the same budget and
same strict preferences should purchase the same object type.

I now present some related results. First, an immediate implication of Theorem \ref{thm:Main} is the
following corollary:
\begin{cor}
  \label{cor:equals}
  Fix an economy $\left(A,\mathcal{A},H,e,x\right)$. The economy is rationalizable only if: whenever
  agents $ik,ik'$ are the same type and $e_{ik}=e_{ik'}$, $x_{ik}=x_{ik'}$.
\end{cor}
\begin{proof}
  Appendix.
\end{proof}
That is, identical agents (of same type and same endowment) must receive the same object type.
Briefly, the theorem requires equal treatment of equals. When types determine both preferences and
endowments, this corollary is an equivalent condition for rationalizability.
\begin{cor}
  Suppose $e_{ik}=e_{ik'}$ for all $k,k'$ and for all $i\in A$. That is, all agents of the same type
  have the same endowment. Then the economy $\left(A,\mathcal{A},H,e,x\right)$ is rationalizable if
  and only if $x_{ik}=x_{ik'}$ for all $k,k'$ and for all $i\in A$. That is, if and only if all
  agents of the same type receive the same object.
\end{cor}
\begin{proof}
  ``Only if'' is a consequence of Corollary \ref{cor:equals}. To prove ``if'', note that everyone of
  the same type receives the same object, so we can let everyone's favorite object be their
  allocated object.
\end{proof}
This resembles the \citet{DS63} theorems for general equilibrium. Their model is an endowment
economy with a finite number of goods, agent types, $k$ copies of each type, and types determining
both endowment and preferences. Only allocations assigning the same bundle to all agents of the same
type are in the core. While neither the \citeauthor{DS63} model nor my model contains the other, it
would be interesting future work to investigate a whether deeper connection exists.

Theorem \ref{thm:Main} characterizes which observed economies are consistent with the core. That is,
the condition offers a restriction on the kinds of allocations that can be seen in equilibrium. Many
allocations can be ruled out ex ante. On the positive side, Corollary \ref{cor:equals} gives a clear
prediction for markets in the core.

Another related question is: what is the minimum number of agent types necessary to rationalize an
allocation? That is, suppose we are free to choose agent types. What is the minimum preference type
heterogeneity required to put $x$ in the core? This question is sensible, since allowing every
individual to be his own type always rationalizes an allocation.

Let $\tilde{\mathcal{A}}$ be the set of individual agents, without encoding information on types.
With this data, we can construct a graph
$\mathcal{\tilde{G}}_{NT}\left(\tilde{\mathcal{A}},H,e,x\right)$ in the same way as
$\mathcal{G}_{NT}\left(A,\mathcal{A},H,e,x\right)$.
\begin{cor}
  Consider $\mathcal{\tilde{G}}_{NT}\left(\mathcal{A},H,e,x\right)$, and decompose this into SCCs,
  $\{S_{1},...,S_{M}\}$. Let $\alpha_{m}$ be the number of distinct object types in $S_{m}$. The
  minimum number of types necessary to construct $\succsim$ such that $x$ is in the core is
  $\alpha=\max\{\alpha_{1},...,\alpha_{m}\}$.
\end{cor}
\begin{proof}
  This is a corollary of Theorem \ref{thm:Main}. Individuals in the same $SCC_{m}$ who receive
  different objects must be different agent types. There is no other restriction on agent types.

  Within $S_{m}$, there are $\alpha_{m}$ distinct objects; order them arbitrarily. Let everyone who
  receives object 1 be type 1, and so on. By Theorem \ref{thm:Main}, this will be rationalizable. It
  is also clear that having fewer than $\alpha$ types will make the economy not NT-rationalizable.
\end{proof}
The result also solves the analogous economy for two-sided matching in the strict core. That is, it
solves a strict stability analogue of \citeauthor{ELSY13} with non-transferable utility. There are
types of men and women, and each type has a strict preference over types of the other side. The
result follows from transforming house-swapping into two-sided matching in the usual way. To do
this, let each agent type have a unique endowment (him- or her- self), and restrict preferences to
find only endowments of the other side acceptable. The condition for rationalizability is given by
Corollary \ref{cor:equals}; an observed market is rationalizable if and only if all men of the same
type are assigned the same type of women, and vice versa.

\subsection{With transfers}

I now turn to the model including monetary transfers. I derive necessary and sufficient conditions
for a T-economy to be T-rationalizable. First, I introduce a new weighted digraph
$\mathcal{G}_{T}\left(A,\mathcal{A},H,(e,\omega),(x,m)\right)=\left(\mathcal{A},E,\ell(\cdot)\right)$.
Draw vertices and arcs as in $\mathcal{G}_{NT}$; let each agent be a node, and draw arcs from $ik$
to \emph{all} vertices $i'k'$ that are endowed with $x_{ik}$. In addition, define the lengths arcs
by $\ell\left(ik,i'k'\right)=\omega_{ik}-m_{ik}$. Note $\ell\left(ik,i'k'\right)$ depends only on
the first vertex, not the second.

The following example adds to \ref{exa:NTexample}.
\begin{example}
  \label{exa:TUexample}
  Consider the economy described in Example \ref{exa:NTexample}, adding the following payments:
  \[
    \begin{array}{cccc}
      \mathcal{A} & e_{ik} & x_{ik} & \omega_{ik}-m_{ik} \\
      1a          & h_{1}  & h_{2}  & 2                  \\
      1b          & h_{2}  & h_{2}  & 0                  \\
      1c          & h_{4}  & h_{5}  & 1                  \\
      2a          & h_{2}  & h_{3}  & -1                 \\
      2b          & h_{5}  & h_{4}  & -1                 \\
      3a          & h_{3}  & h_{1}  & -1
    \end{array}
  \]
  The following figure illustrates $\mathcal{G}_{T}$.

  \begin{figure}[h]
    \centering
    \caption{Figure for Example \ref{exa:TUexample},
      $\mathcal{G}_{T}$}
    \label{fig:ExampleTU-fig}



\begin{tikzpicture}[
roundnode/.style={circle, draw=black, fill=black!5, thick, minimum size=5mm}, 
scale=0.9
]

\tikzstyle{every node}=[font=\footnotesize]

\node[roundnode]    (2a)    at (0,3) 	 	{$2a/h_2$};
\node[roundnode]    (1a)    at (0,0)   		{$1a/h_1$};
\node[roundnode]    (3a)    at (-1.5,1.5) 	{$3a/h_3$};

\node[roundnode]    (1b)    at (3,3)		{$1b/h_2$};

\node[roundnode]    (1c)    at (6,3)		{$1c/h_4$};
\node[roundnode]    (2b)    at (6,0)		{$2b/h_5$};

\path [thick, ->, shorten >=3pt, shorten <=3pt]
    (1a) edge [bend right] node[right] {$2$} (2a) 	
    (2a) edge [bend right] node[above left] {$-1$} (3a) 
    (3a) edge [bend right] node[below left] {$-1$} (1a) 
    (1b) edge [loop right] node[right] {$0$}  (1b) 	
    (1b) edge [bend right] node[above] {$0$}  (2a) 	
    (1a) edge [bend right] node[below right] {$2$}  (1b)
    (1c) edge [bend right] node[left] {$-1$} (2b) 	
    (2b) edge [bend right] node[right] {$1$}  (1c) ;

\end{tikzpicture}

  \end{figure}
\end{example}
We turn to the main result for T-rationalizability.
\begin{thm}
  \label{thm:MainTU}
  Fix a T-economy $\left(A,\mathcal{A},H,(e,\omega),(x,m)\right)$. The following are equivalent:
  \begin{enumerate}
    \item The economy is T-rationalizable.
    \item There exists a vector $p\in\mathbb{R}_{+}^{|H|}$ such that
          \begin{align*}
            (x_{ik}-e_{ik})\cdot p & =\omega_{ik}-m_{ik}\quad\forall ik\in\mathcal{A} & (P)
          \end{align*}
    \item The graph $\mathcal{G}_{T}\left(A,\mathcal{A},H,(e,\omega),(x,m)\right)$ has no cycles
      with length $>0$.
  \end{enumerate}
\end{thm}
\begin{proof}
  Appendix.
\end{proof}
The vector $p$ in $(P)$ (suggestively denoted) is interpretable as a price vector for objects; the
left side is the difference in price between the allocated and endowed object, and the right side is
the net payment from $ik$. This suggests an easy interpretation of the theorem: an economy is
TU-rationalizable if and only if everyone who ``buys'' an object type pays the same price for it.

A direct interpretation of $(3)$ is that no cycles ``export'' money. This is clearly a necessary
condition -- any cycle that pays money outward can implement their object allocation while keeping
more money. In the full proof, I show that $(3)$ also allows construction of a price vector $p$. A
proof sketch follows.
\begin{proof}[Proof sketch of Theorem \ref{thm:MainTU}.]
  I first show $(2)\Longrightarrow(1)$. Given $p$, I seek $v_{i}$ such that $(x,m)$ is a competitive
  equilibrium. By the usual arguments, a competitive equilibrium allocation is in the weak
  core.\footnote{For example, Mas-Colell, Whinston, and Green (1995), pg. 653.} We are looking for
  utility indexes $v_{i}$ such that all agents $ik$ are maximizing subject to their budget
  constraints, given by $e_{ik}\cdot p+\omega_{ik}$. Then this becomes a classic consumer demand
  revealed preference problem. To see this, reinterpret a type $i$ as a single consumer, and each
  individual $ik$ as a demand data point:
  \[
    \{\underset{\text{consumed good and money}}{\underbrace{(x_{ik},m_{ik})}},\underset{\text{budget}}{\underbrace{(e_{ik}\cdot p+\omega_{ik})}},\underset{\text{price}}{\underbrace{p}}\}_{ik}
  \]
  In this structure, such demand data are always rationalizable (in the classical consumer demand
  revealed preference sense). The easiest way to show this is to let $v_{i}(x_{ik})=x_{ik}\cdot p$
  for all $i,ik$, though I show in the full proof this knife-edge construction is not the only one.
  Any utility indexes satisfying Afriat's inequalities work. Then $(x,m)$ is a competitive
  equilibrium supported by $p$, and thus $(x,m)$ is in the weak core.

  I now show $(1)\Longrightarrow(3)$. To see this, note that a cycle $C$'s length
  $\sum_{ik\in C}\omega_{ik}-m_{ik}$ is its members' total net payment of money. If this is greater
  than 0, then this cycle net spends money. Its members can form a blocking coalition -- they can
  allocate objects the same way as in $(x,m)$, but keep their full endowed money for themselves.

  Finally, to show $(3)\Longrightarrow(2)$, I use the shortest path length on $\mathcal{G}_{T}$
  between two objects to construct the price difference between those objects. The construction is
  similar to that in \citet{Quinzii84}. (We can choose a normalization so that $p\geq0$.) In the
  full proof, I show that this construction is consistent -- the minimum path length between objects
  of the same type is always 0. This completes the proof.
\end{proof}
I give an example to illustrate T-rationalizability.
\begin{example*}[Example \ref{exa:TUexample} continued]
  For simplicity, let $\omega_{ik}=3$ for all $ik$. It can be seen that all cycles have length 0, so
  this is rationalizable. Figure \ref{fig:ExampleTU-fig} shows $\mathcal{G}_{T}$, with
  $\omega_{ik}-m_{ik}$ as arc lengths.

  To construct utilities, set $p$ as follows. In the left SCC, let $p_{h_{1}}=3$ arbitrarily, and
  set the prices of other objects in this SCC by the minimum path length from $h_{2}$ plus 3, giving
  $p_{h_{2}}=5,p_{h_{3}}=4$. Notice that the path length between the two copies of $h_{2}$ is 0. In
  the right SCC, let $p_{h_{4}}=1$ arbitrarily, and set $p_{h_{5}}=2$ since the path length from
  $h_{4}$ to $h_{5}$ is 1. Altogether,
  \begin{align*}
    p_{h_{1}} & =3 \\
    p_{h_{2}} & =5 \\
    p_{h_{3}} & =4 \\
    p_{h_{4}} & =1 \\
    p_{h_{5}} & =2
  \end{align*}
  The easiest way to construct T-rationalizing preferences is to let $v_{i}(x_{ik})=x_{ik}\cdot p$
  for all $i$. Though as mentioned above (and demonstrated in the full proof), this is not the only
  construction.
\end{example*}
Theorem \ref{thm:MainTU} establishes a connection between T-rationalizability, competitive
equilibrium, and consumer demand rationalizability. The question of T-rationalizability is
equivalent to consumer demand rationalizability, à la \citeauthor{Samuelson38} and
\citeauthor{Afriat67}. That is, an allocation is rationalizable if and only if each agent type,
interpreted as demand data, is consumer demand rationalizable. Thus, we are looking for utility
indexes such that every agent type is optimizing in their demand. Competitive equilibrium follows.

This yields the theorem's two equivalent and intuitive conditions for T-rationalizability. The first
condition is the existence of a price vector supporting the allocation as a competitive equilibrium.
That is, an allocation is T-rationalizable if and only if it can be supported as a competitive
equilibrium. The second condition is reminiscent of cyclic monotonicity results common in revealed
preference literature. It is readily interpretable directly; a cycle having positive length means it
net transfers money outwards. Then its members can implement the same object allocation while
retaining its money, establishing a blocking coalition.

I now present some corollaries of Theorem \ref{thm:MainTU}. First, I give conditions for strict
T-rationalizability.
\begin{cor}
  \label{cor:strictTUrat}
  Fix a T-economy $(A,\mathcal{A},H,x,m,e,\omega)$. The economy is strictly T-rationalizable if and
  only if \textbf{both} of the following are true:
  \begin{enumerate}
    \item The economy is T-rationalizable.
    \item If $ik,ik'\in S$ are the same type and in the same SCC in $\mathcal{G}_{T}$, then either
      $x_{ik}=x_{ik'}$ OR the shortest path length from $x_{ik}$ to $x_{ik'}$ $\not=0$.
  \end{enumerate}
\end{cor}
\begin{proof}
  Appendix.
\end{proof}
This is the T-rationalizability analogue to Theorem \ref{thm:Main}. The additional condition says
that two individuals of the same type in the same SCC should either be allocated the same object or
pay different amounts. Having a zero path length between $x_{ik}$ and $x_{ik'}$ means their prices
must be the same. Then if two different individuals of type $i$ purchase each one in competitive
equilibrium, they must have the same utility. Conversely, having a nonzero path length allows us to
construct different prices, and thus different utilities.

The following examples illustrate the corollary.
\begin{example*}[Example \ref{exa:TUexample} continued.]
  This example is strictly T-rationalizable. The only thing to check is $x_{1a}$ and $x_{1b}$. Since
  $x_{1a}=x_{1b}$, the economy is strictly TU-rationalizable -- indeed, the utility given in the
  original example suffices.
\end{example*}
\begin{example}
  Suppose instead $x_{1b}=e_{2a}=h_{6}$, a new object type, with no other changes. Focusing on the
  left SCC:
  \[
    \begin{array}{cccc}
      ik & e_{ik} & x_{ik} & \omega_{ik}-m_{ik} \\
      1a & h_{1}  & h_{2}  & 2                  \\
      1b & h_{2}  & h_{6}  & 0                  \\
      2a & h_{6}  & h_{3}  & -1                 \\
      3a & h_{3}  & h_{1}  & -1
    \end{array}
  \]
  \begin{figure}[h]
    \centering
    \caption{Figure for Example \ref{exa:TUexample}
      continued.}
    \label{fig:ExampleTU-nonstrict}



\begin{tikzpicture}[
roundnode/.style={circle, draw=black, fill=black!5, thick, minimum size=5mm}, 
scale=0.9
]

\tikzstyle{every node}=[font=\footnotesize]

\node[roundnode]    (2a)    at (0,3) 	 	{$2a/h_6$};
\node[roundnode]    (1a)    at (0,0)   		{$1a/h_1$};
\node[roundnode]    (3a)    at (-1.5,1.5) 	{$3a/h_3$};

\node[roundnode]    (1b)    at (3,3)		{$1b/h_2$};

\path [thick, ->, shorten >=3pt, shorten <=3pt]
    (1a) edge [bend right] node[below right] {$2$} (1b) 	
    (2a) edge [bend right] node[above left] {$-1$} (3a) 
    (3a) edge [bend right] node[below left] {$-1$} (1a) 
    (1b) edge [bend right] node[above] {$0$}  (2a) ; 	

\end{tikzpicture}

  \end{figure}
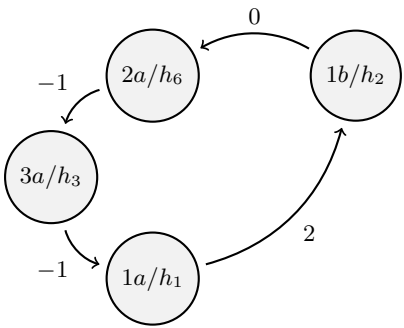
  This economy is T-rationalizable, but not strictly T-rationalizable. The minimum path length from
  $x_{1a}=h_{2}$ to $x_{1b}=h_{6}$ is 0, forcing $p_{h_{2}}=p_{h_{6}}$. If
  $v_{1}(h_{2})>v_{1}(h_{6})$, then $1b$ is not maximizing subject to his budget, so the allocation
  is not a competitive equilibrium and not in the weak core.
\end{example}
The next corollary characterizes possible utility indexes $v_{i}(\cdot)$ that rationalize a
T-economy.
\begin{cor}
  \label{cor:Afriat}
  Fix a T-economy $\left(A,\mathcal{A},H,(e,\omega),(x,m)\right)$. A T-rationalizable economy's
  solutions $v_{i}(\cdot)$ are characterized by solutions to the following linear system.
  \begin{align*}
    \text{for \ensuremath{p} s.t. } & (x_{ik}-e_{ik})\cdot p=\omega_{ik}-m_{ik}\quad\forall ik\in\mathcal{A}:                                                                                                                      \\
    1.\                             & v_{i}(x_{ik})\leq v_{i}(x_{ik'})+p\cdot(x_{ik}-x_{ik'})-w_{ik'}\quad\forall i,\forall ik,ik'                                                                                                 \\
    2.\                             & \text{for any \ensuremath{h} such that \ensuremath{h\not=x_{ik}} \ensuremath{\forall x_{ik}} , and for any \ensuremath{ik} such that \ensuremath{h\cdot p\leq e_{ik}\cdot p+\omega_{ik}} : } \\
                                    & \quad\quad v_{i}(h)-h\cdot p\leq v_{i}(x_{ik})-x_{ik}\cdot p
  \end{align*}
\end{cor}
\begin{proof}
  Appendix.
\end{proof}
The first line characterizes valid price vectors $p$. The inequalities define the utility indexes
$\left(v_{i}\right)$ given a valid $p$. Inequality $1.$ is the Afriat inequality for quasilinear
utility (with marginal utility of money equal to one). Inequality $2.$ gives restrictions on
utilities for any objects that are never consumed by type $i$. If an object $h$ is never consumed
but is affordable under some budget $e_{ik}\cdot p+\omega_{ik}:=I_{ik}$, its consumption bundle
$(h,I_{k}-h\cdot p)$ must be dispreferred to the actually consumed bundle
$(x_{ik},I_{k}-x_{ik}\cdot p)$.

This linear system gives possible values of $(v_{i})$ from the observed data. As is the case in
consumer demand, these are joint restrictions rather than valid ranges for each $v_{i}(h)$. For
example, there are many possible price vectors, each leading to a range of possible utility indexes
$v_{i}$'s. I also show in the proof of Theorem \ref{thm:MainTU} that relative prices are determined
within an SCC but not across SCCs.\footnote{For this reason I conjecture it is not possible to write
a linear system without the existential statement of $(P)$.} Nevertheless, Corollary
\ref{cor:Afriat} characterizes the joint restrictions for valid $v_{i}$s.

\section{Conclusion}

In this paper, I present testable implications of the core in exchange economies with and without
monetary transfers. The key identifying assumption is on agent types -- that preferences are solely
a function of observable characteristics of the agents. The analyst observes these types,
endowments, and allocations, but not the preferences. Given this, I derive tractable and intuitive
conditions for the core to be rationalizable.

The conditions in Theorems \ref{thm:Main} and \ref{thm:MainTU} characterize markets that are
compatible with the core. That is, they can falsify a market being in the core; they also serve as
ex ante predictions for market outcomes. The results can also be applied to audit mechanisms when
the matching procedure is unknown.

The work here suggests paths for future research. One takeaway is that other information must be
observed (such as partial data on preferences or some structure of the allocation process) to
further distinguish outcomes. Analogously to the development of GARP, one can implement smoother
measures of rationalizability or construct statistical tests for rationalizability. The tractability
of the graph construction $\mathcal{G}_{NT}$ and $\mathcal{G}_{T}$ may be useful in such work.

\pagebreak

\appendix

\section{Results for \texorpdfstring{$\mathcal{G}_{NT}$}{G\_NT}}

First, I introduce another graph construction. Given a NT-economy\footnote{Or, if given a T-economy,
discard $\omega$ and $m$.}, draw
$\mathcal{G}_{small}\left(A,\mathcal{A},H,e,x\right)=(\mathcal{A},E')$ as follows:
\begin{enumerate}
  \item[Initialize.] Draw all agents $\mathcal{A}$ as vertices.
  \item[Step $m$.] Consider all agents receiving $h_{m}$, that is all $ik$ such that $x_{ik}=h_{m}$.
    Order them according to their index; refer to these as the ``left'' side. Similarly, order
    agents endowed with $h_{m}$ according to their index; these are the ``right'' side. By
    construction, these two sets are the same cardinality. Draw one arc from the first agent on the
    left side to the first agent on the right side, and so on. If $m<\eta$, continue to step $m+1$.
\end{enumerate}
The graph produced after $|H|$ steps represents the allocation $\mu$. Note that each agent has one
out-arc and one in-arc. Recall the construction of $\mathcal{G}_{NT}=(\mathcal{A},E)$. Note also
that $E\supseteq E'$; that is, $\mathcal{G}_{NT}$ can be obtained by adding arcs to
$\mathcal{G}_{small}$. Figure \ref{fig:Example-BothFigs} shows both constructions for Example
\ref{exa:NTexample}.

\begin{figure}[h]
  \centering
  \caption{Figure for Example \ref{exa:NTexample}}
  \label{fig:Example-BothFigs}
  \begin{minipage}[t]{0.4\columnwidth}%
    \begin{center}
      Graph $\mathcal{G}_{NT}$
    \end{center}
    \begin{center}

\begin{tikzpicture}[
roundnode/.style={circle, draw=black, fill=black!5, thick, minimum size=5mm}, 
scale=0.7
]

\tikzstyle{every node}=[font=\scriptsize]

\node[roundnode]    (2a)    at (0,3) 	 	{$2a/h_2$};
\node[roundnode]    (1a)    at (0,0)   		{$1a/h_1$};
\node[roundnode]    (3a)    at (-1.5,1.5) 	{$3a/h_3$};

\node[roundnode]    (1b)    at (3,3)		{$1b/h_2$};

\node[roundnode]    (1c)    at (6,3)		{$1c/h_4$};
\node[roundnode]    (2b)    at (6,0)		{$2b/h_5$};

\path [thick, ->, shorten >=3pt, shorten <=3pt]
    (1a) edge [bend right] (2a) 	
    (2a) edge [bend right] (3a) 
    (3a) edge [bend right] (1a) 
    (1b) edge [loop right] (1b) 	
    (1b) edge [bend right] (2a) 	
    (1a) edge [bend right] (1b)
    (1c) edge [bend right] (2b) 	
    (2b) edge [bend right] (1c) ;

\end{tikzpicture}
    \end{center}%
  \end{minipage}\hfill%
  \begin{minipage}[t]{0.4\columnwidth}%
    \begin{center}
      (A particular) Graph $\mathcal{G}_{small}$
    \end{center}
    \begin{center}

\begin{tikzpicture}[
roundnode/.style={circle, draw=black, fill=black!5, thick, minimum size=5mm}, 
scale=0.7
]

\tikzstyle{every node}=[font=\scriptsize]

\node[roundnode]    (2a)    at (0,3) 	 	{$2a/h_2$};
\node[roundnode]    (1a)    at (0,0)   		{$1a/h_1$};
\node[roundnode]    (3a)    at (-1.5,1.5) 	{$3a/h_3$};

\node[roundnode]    (1b)    at (3,3)		{$1b/h_2$};

\node[roundnode]    (1c)    at (6,3)		{$1c/h_4$};
\node[roundnode]    (2b)    at (6,0)		{$2b/h_5$};

\path [thick, ->, shorten >=3pt, shorten <=3pt]
    (1a) edge [bend right] (1b) 	
    (1b) edge [bend right] (2a)
    (2a) edge [bend right] (3a) 
    (3a) edge [bend right] (1a) 
    (1c) edge [bend right] (2b) 	
    (2b) edge [bend right] (1c) ;

\end{tikzpicture}
    \end{center}%
  \end{minipage}
\end{figure}

\begin{figure}[H]
  \centering
  \caption{Illustration of Proposition \ref{prop:scc-supercycle}}
  \label{fig:supercycle}
  \begin{minipage}[t]{0.4\columnwidth}%
    \begin{center}
      Standard case
    \end{center}
    \begin{center}

\begin{tikzpicture}[
roundnode/.style={circle, draw=black, fill=black!5, thick, minimum size=8mm}, 
scale=1
]

\node[roundnode]    (1a)    at (0,3)    {$1a$};
\node[roundnode]    (2a)    at (0,0)    {$2a$};

\node[roundnode]    (3a)    at (3,3)    {$3a$};
\node[roundnode]    (4a)    at (3,0)    {$4a$};

\path [thick, ->, shorten >=3pt, shorten <=3pt]
    (2a) edge [bend right] (1a) 
    (3a) edge [bend right] (4a) 
    (2a) edge [bend right, dashed] (4a)
    (3a) edge [bend right, dashed] (1a) ;


\draw [->, dotted, thick] (1a) to [out=150, in=-150, looseness=2] node[below, sloped] {rest of big-cycle} (2a);
\draw [->, dotted, thick] (4a) to [out=-30, in=30, looseness=2] node[below, sloped] {rest of small-cycle} (3a);

\end{tikzpicture}

    \end{center}%
  \end{minipage}\hfill%
  \begin{minipage}[t]{0.4\columnwidth}%
    \begin{center}
      Self-loop
    \end{center}
    \begin{center}

\begin{tikzpicture}[
roundnode/.style={circle, draw=black, fill=black!5, thick, minimum size=8mm},
scale=1
]

\node[roundnode]    (1a)    at (0,3)    {$1a$};
\node[roundnode]    (2a)    at (0,0)    {$2a$};

\node[roundnode]    (3a)    at (3,3)    {$3a$};

\path [thick, ->, shorten >=3pt, shorten <=3pt]
    (2a) edge [bend right] (1a) 
    (2a) edge [bend right, dashed] (3a) 
    (3a) edge [bend right, dashed] (1a) 
    (3a) edge [loop, looseness=5, out=-30, in=30] (3a);

\draw [->, dotted, thick] (1a) to [out=150, in=-150, looseness=2] node[below, sloped] {rest of big-cycle} (2a);

\end{tikzpicture}

    \end{center}%
  \end{minipage}
\end{figure}
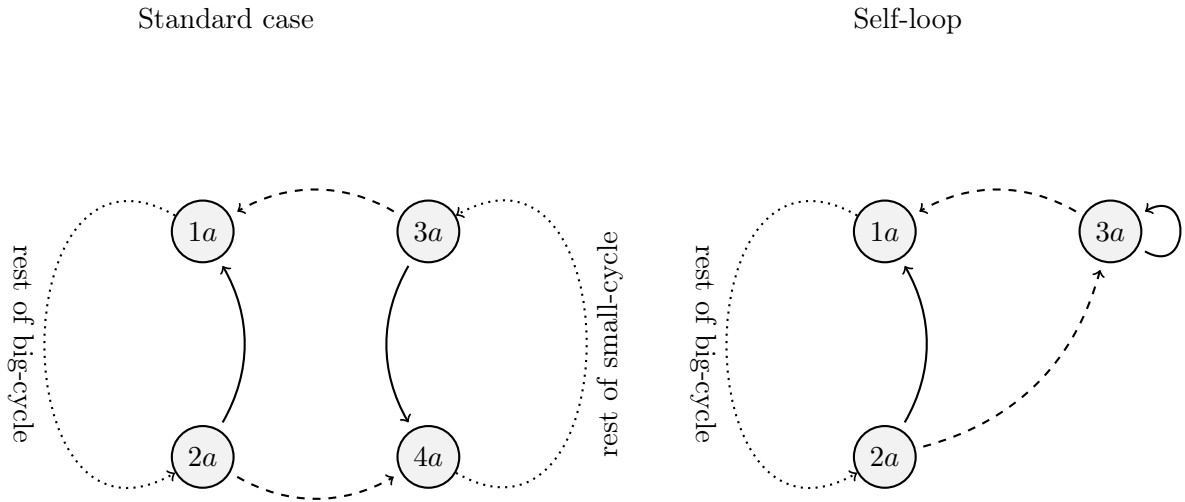

I now provide some intermediate results related to the constructed graphs $\mathcal{G}_{small}$ and
$\mathcal{G}_{NT}$. These will be key for the proof of Theorem \ref{thm:Main}.
\begin{prop}
  \label{prop:small-cycles}
  Consider $\mathcal{G}_{small}\left(A,\mathcal{A},H,e,x\right)=(\mathcal{A},E')$.
  $\mathcal{G}_{small}$ has a subgraph partition into cycles. That is, there are disjoint subgraphs
  $C_{1},...,C_{N}$ such that $\mathcal{G}_{small}=\cup_{n=1}^{N}C_{n}$, $C_{m}\cap C_{n}=\emptyset$
  for all $m,n$, and each $C_{n}$ is a cycle.
\end{prop}
\begin{proof}
  Note each vertex $i$ has $d^{-}(ik)=d^{+}(ik)=1$. We can invoke a version of Veblen's theorem:
  \begin{quote}
    (Veblen's theorem) A directed graph $D=(V,E)$ admits a partition of arcs into cycles if and only
    if $d^{-}(v)=d^{+}(v)$ for all vertices $v\in V$. \citep{Veblen12,BM08}
  \end{quote}
  Since $d^{-}(ik)=d^{+}(ik)$, $\mathcal{G}_{small}$ has a partition of arcs into cycles. There are
  no isolated vertices, so every vertex is in at least one cycle. Further, since
  $d^{-}(ik)=d^{+}(ik)=1$ each vertex must be in at most one cycle. Thus the arc partition into
  cycles also partitions the vertices into cycles.
\end{proof}
\begin{prop}
  \label{prop:scc-supercycle}
  Consider $\mathcal{G}_{NT}\left(A,\mathcal{A},H,e,x\right)$. For every strongly connected
  component $S$ of $\mathcal{G}_{NT}$, there is a cycle including all vertices in $S$.
\end{prop}
\begin{proof}
  By Proposition \ref{prop:small-cycles}, $\mathcal{G}_{small}$ admits a partition of vertices into
  cycles. Recall $\mathcal{G}_{NT}=(\mathcal{A},E)$ and $\mathcal{G}_{small}=(\mathcal{A},E')$,
  where $E\supseteq E'$. Then these cycles also partition $\mathcal{G}_{NT}$'s vertices. The SCC $S$
  in $\mathcal{G}_{NT}$ is composed of the vertices in a number of $\mathcal{G}_{small}$-cycles. It
  cannot include a strict subset of vertices in a $\mathcal{G}_{small}$-cycle since there is always
  a path between any two vertices in a cycle.

  The remaining argument is by strong induction on the number $K$ of $\mathcal{G}_{small}$-cycles
  contained in $S$. Assign an order to these cycles in the following way. Let the first cycle be any
  of these. Choose the $k^{th}$ cycle such that it has the same object type as one of the first
  $k-1$ cycles. It is always possible to do this -- suppose at some point none of the remaining
  cycles has the same object type as the first $k$ cycles. Then there are no paths in
  $\mathcal{G}_{NT}$ between the first $k$ cycles and the remaining cycles (recall arcs are drawn
  from an agent to all agents whose endowment he receives), so they are not in the same SCC.
  \begin{enumerate}
    \item[Claim.] There is a cycle in $\mathcal{G}_{NT}$ covering all vertices in the first $k$
      $\mathcal{G}_{small}$-cycles in $S$. As shorthand, I will call this the ``big-cycle'', and the
      $\mathcal{G}_{small}$-cycles will be ``small-cycles''.
    \item[Base claim.] For $k=1$, the claim is trivial.
    \item[$k^{th}$ claim.] Suppose the claim is true for the first $k-1$ cycles. That is, there is a
      $k-1^{th}$ big-cycle in $\mathcal{G}_{NT}$ covering all the vertices in the first $k-1$
      small-cycles. I show that there is a cycle covering all vertices in the $k-1^{th}$ big-cycle
      and the $k^{th}$ small-cycle. The following argument is illustrated in Figure
      \ref{fig:supercycle}. There are three cases, depending on whether either cycle is a self-loop.
          \begin{enumerate}
            \item[Case 1.] Suppose neither is a self-loop. Let the big-cycle be $(1a,...,2a,1a)$,
              and the $k^{th}$ small-cycle be $(3a,4a,...,3a)$. That is, $x_{2a}=e_{1a}$ and so on.
              I do not require that the denoted agents are all different types; e.g. $2a$ can be
              $1b$. By the ordering of the cycles, the $k^{th}$ small-cycle and the $k-1^{th}$
              big-cycle have at least one of the same object type. Without loss of generality let
              $e_{1a}=e_{4a}$. This gives $x_{2a}=e_{1a}=e_{4a}$, so we have the arc $(2a,4a)\in E$.
              Similarly, $x_{3a}=e_{4a}=e_{1a}$, so we have the arc $(3a,1a)\in E$. This gives us a
              new big-cycle across all the vertices in the first $k$ small-cycles:
              $(\underset{\text{big-cycle }k-1}{\underbrace{1a,...,2a}},\underset{k^{th}\text{ cycle}}{\underbrace{4a,...,3a}},1a)$.
            \item[Case 2.] Suppose the $k^{th}$ small-cycle is a self-loop, but the $k-1^{th}$
              big-cycle is not. Then let the big-cycle be $(1a,...,2a,1a)$, and the $k^{th}$
              small-cycle be $(3a,3a)$. Again, let $e_{1a}=e_{3a}$ without loss of generality. Then
              $x_{2a}=e_{1a}=e_{3a}$ implies $(2a,3a)\in E$. Likewise, $x_{3a}=e_{3a}=e_{1a}$
              implies $(3a,1a)\in E$. So we have a new big-cycle
              $(\underset{\text{big-cycle }k-1}{\underbrace{1a,...,2a}},3a,1a)$. The case if the
              big-cycle is a self-loop is the same (this may occur in the $k=2$ claim).
            \item[Case 3.] Suppose both are self-loops. Then let the big-cycle be $(1a,1a)$ and the
              $k^{th}$ small-cycle be $(3a,3a)$. Again, we suppose $e_{1a}=e_{3a}$. Then
              $x_{1a}=e_{1a}=e_{3a}$ implies $(1a,3a)\in E$, and likewise $(3a,1a)\in E$. So we have
              a new big-cycle $(1a,3a,1a)$.
          \end{enumerate}
  \end{enumerate}
  This completes the proof.
\end{proof}

The following lemma is derived from Proposition \ref{prop:scc-supercycle} and its proof.
\begin{lem}
  \label{lem:no-inout}
  Consider $\mathcal{G}_{NT}\left(A,\mathcal{A},H,e,x\right)$. Every strongly connected component
  $S$ has no in- or out- arcs. That is, if $ik\in S$ and $(ik,i'k')\in E$ or $(i'k',ik)\in E$, then
  $i'k'\in S$. Alternatively, the strongly connected components and (weak) components coincide.
\end{lem}
\begin{proof}
  There is a cycle covering all vertices of $S$ by Proposition \ref{prop:scc-supercycle}. Suppose
  there is an out-arc from $S$ pointing to a vertex in a different SCC $S'$. $S'$ also has a cycle
  covering all its vertices. The same argument as in the induction part of the proof of Proposition
  \ref{prop:scc-supercycle} establishes an arc from $S'$ to $S$. Thus there are paths from between
  any vertices in $S$ and $S'$, and they are in the same SCC, a contradiction. The case for no
  in-arcs is a relabeling of $S$ and $S'$.
\end{proof}
The following is a corollary of Lemma \ref{lem:no-inout}.
\begin{cor}
  \label{cor:path-equivalence}
  Consider $\mathcal{G}_{NT}\left(A,\mathcal{A},H,e,x\right)$. Let $ik$ and $i'k'$ be distinct
  vertices. There exists a $(ik,i'k')$-path if and only if $ik$ and $i'k'$ are in the same SCC.
  Equivalently, there exists a $(ik,i'k')$-path if and only if there exists a $(i'k',ik)$-path.
\end{cor}
\begin{proof}
  If $ik$ and $i'k'$ are in the same SCC, there exists a $(ik,i'k')$-path by definition. Suppose
  there exists a $(ik,i'k')$-path. By Lemma \ref{lem:no-inout}, there are no paths between different
  SCCs, so $ik$ and $i'k'$ must be in the same SCC.
\end{proof}
\begin{cor}
  \label{cor:SCCalltypes}
  Consider $\mathcal{G}_{NT}\left(A,\mathcal{A},H,e,x\right)$. All copies of the same object type
  are in the same SCC. That is, if $e_{ik}=e_{i'k'}$ and $ik\in S$, then $i'k'\in S$.
\end{cor}
\begin{proof}
  Let $e_{ik}=e_{i'k'}$. There is at least one agent pointing to $ik$, so $\exists a\in\mathcal{A}$
  such that $(a,ik)\in E$. Then $(a,i'k')\in E$ as well by construction. By Corollary
  \ref{cor:path-equivalence}, there are $(ik,a)$- and $(i'k',a)$- paths. Then there are $(ik,i'k')$-
  and $(i'k',ik)$- paths (through $a$), so $ik$ and $i'k'$ are in the same SCC.
\end{proof}
The above results give us significant information about the SCCs of $\mathcal{G}_{NT}$. The
following is a summary of these results. From Proposition \ref{prop:scc-supercycle}, each SCC
contains a cycle covering all its vertices. From Lemma \ref{lem:no-inout} and Corollary
\ref{cor:path-equivalence}, $\mathcal{G}_{NT}$ can be vertex- \emph{and }arc- partitioned into its
SCCs. That is, $\mathcal{G}_{NT}$ consists of SCCs with no links between them. Finally, Corollary
\ref{cor:SCCalltypes} tells us all copies of a given object type are in the same SCC.

If we take Theorem \ref{thm:Main} as given for now, we can use the above result to prove Corollary
\ref{cor:equals}.
\begin{proof}[Proof of Corollary \ref{cor:equals}]
  If if $e_{ik}=e_{ik'}$, then $ik$ and $ik'$ are in the same SCC. Then apply Theorem \ref{thm:Main}
  to get the desired result.
\end{proof}

\section{Proof of Theorem \ref{thm:Main} }
\begin{proof}[Proof of Theorem \ref{thm:Main}]
  (``If'') Let the supposition be true: whenever agents of the same type are in the same SCC, they
  receive the same object type. I find a preference profile $\succsim$ that such that $x$ is in the
  core. First find the partition of vertices into SCCs. Then assign an arbitrary order to the SCCs,
  and denote them $S_{1},...S_{M}$. Construct the preferences by the following procedure. Let
  $\succsim_{i}(n)$ denote type $i$'s $n^{th}$ favorite object.
  \begin{enumerate}
    \item[Step 1.] In $S_{1}$, for all $i\in S_{1}$, let $\succsim_{i}(1)=x_{i}$. This is well
      defined since if there are multiple agents of the same type in $S_{1}$, they all receive the
      same object type.
    \item[Step 2.] In $S_{2}$, for all $i\in S_{2}$, let $\succsim_{i}(1)=x_{i}$ if possible. This
      is possible if there were no type $i$'s in $S_{1}$. Otherwise, let $\succsim_{i}(2)=x_{i}$. By
      Corollary \ref{cor:SCCalltypes}, an object never reappears in a later step, so this never
      assigns an object to two places in the same preference.
    \item[Step $m$.] In $S_{m}$ for $m=2,...,M$, for all $i\in S_{k}$, let $\succsim_{i}(m')=x_{i}$
      for the lowest unassigned $m'=2,...,m$. Again by the same argument above, this never assigns
      two objects to the same type; it also never assigns the same object type to multiple places in
      the same preference.
    \item[Step $M+1$.] Assign remaining preferences in any order, if necessary.
  \end{enumerate}
  I now show this preference profile admits no blocking coalition. Suppose that there is a coalition
  of agents $A'\subseteq\mathcal{A}$ and feasible sub-allocation $\mu'$ such that for all
  $ik\in A':x'_{ik}\succsim_{i}x_{ik}$. The argument is by induction on the number of SCCs $M$. In
  each SCC $S_{m}$, the claim to demonstrate is that $x_{ik}'=x_{ik}$ for all $ik\in A'\cap S_{m}$.
  \begin{enumerate}
    \item[Base case.] In $S_{1}$, all agents receive their favorite object. Then
      $x'_{ik}\sim x_{ik}$ for all $i\in A'\cap S_{1}$. The only indifferences are between copies of
      the same object type, so this gives $x'_{ik}=x_{ik}$.
    \item[$m^{th}$ case.] Suppose the claim is true for all agents in
      $A'\cap\left(S_{1}\cup\cdots\cup S_{m-1}\right)$. This implies that $x'$ allocates all agents
      in $A'\cap\left(S_{1}\cup\cdots\cup S_{m-1}\right)$ objects in their own SCC. That is,
      $x'_{ik}\in\cup_{ik\in A'\cap S_{m}}e_{ik}$.

          Toward a contradiction, suppose that $\exists ik\in S_{m}$ such that
          $x_{ik}':=h\succ_{i}x_{ik}$. Then it must be
          $h\in\cup_{ik\in S_{1}\cup\cdots\cup S_{m-1}}e_{ik}$, since all strictly preferred objects
          are in earlier SCCs. Further, since $x'$ reallocates within $A'$, it must be
          $h\in\cup_{ik\in A'\cap\left(S_{1}\cup\cdots\cup S_{m-1}\right)}e_{ik}$. Then it must be
          that an agent in $A'\cap\left(S_{1}\cup\cdots\cup S_{m-1}\right)$ receives an object in
          $\cup_{ik\in A'\cap\left(S_{1}\cup\cdots\cup S_{m-1}\right)}e_{ik}$. This contradicts the
          supposition, so it must be that $x_{ik}'\sim x_{ik}$ for $ik\in A'\cap S_{m}$, giving
          $x'_{ik}=x_{ik}$.

  \end{enumerate}
  Thus $x_{ik}'=x_{ik}$ for all $ik\in A'$, and $A'$ is not a blocking coalition.

  (``Only if'') Toward the contrapositive, suppose there is an SCC $S$ with two agents of the same
  type who receive different objects. By Proposition \ref{prop:scc-supercycle}, there is a cycle
  covering all vertices in $S$. I now construct a blocking coalition using this cycle. Note that two
  of these vertices represent agents of the same type who receive different objects. Let these two
  agents be $1a$ and $1b$; I consider cases based on their relative positions in the cycle.
  \begin{enumerate}
    \item Suppose the cycle is
      $1a\rightarrow\underset{:=c}{\underbrace{2a\rightarrow\cdots\rightarrow1b}}\rightarrow3a\rightarrow\cdots\rightarrow1a$,
      and $e_{2a}\not=e_{3a}$. Suppose $e_{2a}\succ_{1}e_{3a}$. Then
      $1b\rightarrow\underset{c}{\underbrace{2a\rightarrow\cdots\rightarrow1b}}$ represents a
      blocking coalition. Note that this is a feasible sub-allocation; it contains its own
      endowment, and $1b$ is strictly better off. The case $e_{2a}\prec_{1}e_{3a}$ is a rotation and
      relabeling of the cycle.
    \item Suppose the cycle is
      $1a\rightarrow1b\rightarrow\underset{:=c}{\underbrace{2a\rightarrow\cdots\rightarrow1a}}$. If
      $e_{2a}\succ_{1}e_{1b}$, then
      $1a\rightarrow\underset{c}{\underbrace{2a\rightarrow\cdots\rightarrow1a}}$ is a blocking
      coalition. If instead $e_{1b}\succ_{1}e_{2a}$, then $x$ is not individually rational for $1b$.
    \item If the cycle is $1a\rightarrow1b\rightarrow1a$ and $e_{1a}\not=e_{1b}$, then $\mu$ is not
      individually rational.
  \end{enumerate}
  This completes the proof.
\end{proof}
\begin{rem*}
  For readers familiar with the result in \citet{QW04}, it suffices to show that executing their
  ``$\mathcal{STRICTCORE}$'' algorithm on the above constructed preferences results in the
  allocation $\mu$. This is straight forward, and a formal proof is omitted.
\end{rem*}

\section{Proof of Theorem \ref{thm:MainTU} and related results}

I first give a formal definition of competitive equilibrium in an exchange economy setting.
\begin{defn}
  \label{def:CE}
  Let $E=\left\{ (\omega_{ik},e_{ik}),(u_{ik})\right\} _{ik\in\mathcal{A}}$ be an exchange economy,
  where $u_{ik}(\cdot):H\times\mathbb{R}_{+}\rightarrow\mathbb{R}$ are utility functions. A
  \textbf{competitive equilibrium} is a price vector $p\in\mathbb{R}^{H}$ and a feasible allocation
  $(x_{ik},m_{ik})_{ik\in\mathcal{A}}$ such that for all $ik\in\mathcal{A}$:
  \begin{itemize}
    \item $m_{ik}+p\cdot x_{ik}\leq\omega_{ik}+p\cdot e_{ik}$
    \item $\left(u_{ik}(h,m)>u_{ik}(x_{ik},m_{ik})\right)\Longrightarrow\left(m+p\cdot h>\omega_{ik}+p\cdot e_{ik}\right)$
  \end{itemize}
  That is, all agents' allocations are affordable for them, and any better allocation is
  unaffordable. A \textbf{competitive equilibrium allocation} is
  $(x_{ik},m_{ik})_{ik\in\mathcal{A}}$ for which there exists a price vector supporting it as a
  competitive equilibrium.
\end{defn}
\begin{defn}
  Let $\left\{ (x^{r},p^{r},I^{r})\right\} _{r=1}^{N}$ be observed demand, price, and budget data,
  where $x^{r}\in\mathbb{R}_{+}^{H};p^{r},I^{r}\in\mathbb{R}_{++}^{H}$. The data is
  \textbf{quasilinear rationalizable} if for all $r$, $x^{r}$ solves
  \begin{align*}
    \max_{x\in\mathbb{R}_{++}^{n}}v(x)+m \\
    \text{s.t. }p^{r}\cdot x+m=I^{r}
  \end{align*}
  for some concave $v$.
\end{defn}
I also give Afriat's theorem for quasilinear rationalizability. (These are the usual Afriat
inequalities with $\lambda=1$.)
\begin{thm}[Afriat's theorem]
  \label{thm:Afriat}
  Data $(x^{r},p^{r},I^{r}),r=1,...,N$ are quasilinear rationalizable if and only if there exist
  numbers $v_{j}\in\mathbb{R}$ such that
  \begin{align*}
    v^{k} & \leq v^{j}+\left(p^{j}\cdot x^{r}-I^{r}\right) & (A)
  \end{align*}
\end{thm}
I now give the full proof for Theorem \ref{thm:MainTU}.
\begin{proof}[Proof of Theorem \ref{thm:MainTU}.]
  I show $(2)\Rightarrow(1)\Rightarrow(3)\Rightarrow(2)$.

  $(2)\Longrightarrow(1)$. Suppose there exists a vector $p$ satisfying equation $(P)$. I first seek
  to show that this $p$ supports $(x,m)$ as a competitive equilibrium for some utility indexes
  $(v_{i})$. That is, I want to construct $v_{i}$ such that all agents $ik$ are maximizing utility
  subject to their budget constraints $e_{ik}'\cdot p+\omega_{ik}$.\footnote{Agent $ik$ sells his
  endowment $e'_{ik}$ at price $p$ and is additionally endowed with $\omega_{ik}$ money.} This
  becomes a classic consumer demand revealed preference problem. To see this, reinterpret an agent
  type $i$ as a single consumer, and each individual agent $ik$ as a demand data point from this
  consumer:
  \[
    \left(\underset{\text{consumed good and money}}{\underbrace{(x_{ik},m_{ik})}},\underset{\text{budget}}{\underbrace{(e_{ik}'\cdot p+\omega_{ik}):=I_{ik}}},\underset{\text{price}}{\underbrace{p}}\right)_{k\in\{1,...,K_{i}\}}
  \]
  That is, $i$ is a consumer, and each $ik$ is a single observation of demand at a particular
  budget. There are $|A|$ consumers and $K_{i}$ demand points for each consumer $i$. We seek to
  rationalize the demand data in a consumer revealed demand sense by constructing $(v_{i})$ such
  that each consumer $i$ is maximizing utility $V_{i}(h,m)=v_{i}(h)+m$ in each consumption
  bundle-budget pair.

  The easiest way to do this is to let $v_{i}(x_{ik})=x_{ik}'\cdot p$, making all agents indifferent
  to any possible consumption bundle while still satisfying assumption $(A2)$. However, these data
  are rationalizable more broadly; any indexes fulfilling Afriat's inequalities $(A)$ will also
  suffice for $(v_{i})$.

  $(1)\Longrightarrow(3)$. I show the contrapositive; suppose $\mathcal{G}_{T}$ has a cycle $C$ with
  positive length; i.e. $\sum_{ik\in C}\omega_{ik}-m_{ik}>0$. The members of $C$ can form a blocking
  coalition for $(x,m)$ by allocating to each $ik\in C$
  \[
    \left(x_{ik},m+\frac{\sum_{ik\in C}\omega_{ik}-m_{ik}}{|C|}\right)
  \]
  That is, each agent receives the same object and receives more money from the excess endowment.
  This is feasible for $C$ and strictly preferred by all $ik\in C$.

  $(3)\Longrightarrow(2)$. Suppose $\mathcal{G}_{T}$ has no cycles with length $>0$. I construct a
  price $p$ satisfying $(P)$ via path lengths on $\mathcal{G}_{T}$. Note that Proposition
  \ref{prop:scc-supercycle}, Lemma \ref{lem:no-inout}, and Corollary \ref{cor:SCCalltypes} still
  apply to $\mathcal{G}_{T}$. Every SCC has a cycle covering all its vertices; there are no paths
  between two SCCs; and all objects of the same type are in the same SCC. Denote $p_{h}$ as the
  price of object type $h\in H$. Construct $p$ as follows:
  \begin{enumerate}
    \item For each SCC, choose any object type $h$ in this SCC and set $p_{h}$ to be any number.
    \item For all objects $h'$ in this SCC, set $p_{h'}-p_{h}$ to be length of the shortest path
      from $h$ to $h'$. That is, the shortest path between an agent endowed with $h$ to an agent
      endowed with $h'$ determines the price difference.
    \item Repeat steps 1 and 2 for all SCCs.
    \item Add a constant to $p$ to ensure $p\geq0$.
  \end{enumerate}
  I will show that all paths between two vertices are the same length, then that the path length
  between an object type $h$ and itself is always 0, so that the construction is consistent, i.e.
  $p_{h}-p_{h'}=0$ when $h=h'$. The rest of the proof will immediately follow.

  Note the whole economy is budget balanced; we have
  $\sum_{ik\in\mathcal{A}}\omega_{ik}=\sum_{ik\in A}m_{ik}$. For any cycles that form a
  vertex-partition of $\mathcal{G}_{T}$: these cycles must have length 0. A negative length cycle
  that is in a partition of the overall economy implies a positive length cycle elsewhere by budget
  balancedness, a contradiction.

  In particular, by Proposition \ref{prop:scc-supercycle}, each SCC has a cycle containing all its
  vertices; call this the ``whole-cycle'' as shorthand. These partition the whole economy, so each
  whole-cycle must have length 0. For the following claims, assume the SCC has at least three
  vertices. I will show the cases for one or two vertices separately. Enumerate the whole-cycle as
  $(1a,2a,...,sa,...(S-1)a,Sa,1a)$. (Allowing any of these agents to be of the same type -- this is
  unimportant.) Now consider $1a$ and $sa$ distinct and in the same SCC (recall there are no paths
  between SCCs), and consider the path $(1a,...,sa)$ via the whole-cycle. Denote this path
  $(1a,\underset{:=\alpha}{\underbrace{2a,...,\left(s-1\right)a}},sa)$, and call it the
  ``whole-cycle path'' as shorthand.
  \begin{claim}
    \label{claim:samelength}
    If the arc $(1a,sa)$ exists, it is the same length as the whole-cycle path. That is,
    $\ell(1a,sa)=\ell(1a,2a,...,\left(s-1\right)a,sa)$.
  \end{claim}
  Figure \ref{fig:samelength} illustrates the following argument. If the arc $(1a,sa)$ exists, then
  $e_{2a}=e_{sa}$, so there is an arc $((s-1)a,2a)$. Then
  $(\underset{:=\alpha}{\underbrace{2a,...,(s-1)a}},2a)$ forms a cycle, and
  $(1a,sa,\underset{\text{rest of whole-cycle}}{\underbrace{...}},1a)$ also forms a cycle. Since the
  two cycles partition the SCC, they are part of a partition of the overall economy; thus both
  cycles must have length 0. If $\ell(1a,sa)>\ell(1a,2a,...,\left(s-1\right)a,sa)$, then the latter
  cycle has positive length, a contradiction. This is because the whole-cycle has length 0 as
  established, and we have found a cycle with shorter length. If instead
  $\ell(1a,sa)<\ell(1a,2a,...,\left(s-1\right)a,sa)$, then the latter cycle has negative length,
  also a contradiction. Note the same argument carries through if $2a=(s-1)a$ -- the first cycle is
  a self-loop, and $1a=(s-1)a$ is symmetric.

  \begin{figure}[H]
    \centering
    \caption{Illustration of Claim \ref{claim:samelength}}
    \label{fig:samelength}



\begin{tikzpicture}[
roundnode/.style={circle, draw=black, fill=black!5, thick, minimum size=11mm},
scale=0.9
]

\tikzstyle{every node}=[font=\footnotesize]

\node[roundnode]    (1)    at (0,3)    {$(s-1)a$};
\node[roundnode]    (2)    at (0,0)    {$2a$};

\node[roundnode]    (3)    at (2,4)    {$sa$};
\node[roundnode]    (4)    at (2,-1)    {$1a$};

\path [thick, ->, shorten >=3pt, shorten <=3pt]
    (4) edge [] node[below right] {}   (3)
    (4) edge [bend left, dashed] (2)
    (2) edge [bend left, dashed] node[left] {$\alpha$}  (1)
    (1) edge [bend left, dashed] (2)
    (1) edge [bend left] (2)
    (1) edge [bend left, dashed] (3);

\draw [->, dotted, thick] (3) to [out=30, in=-30, looseness=1.5] node [above, sloped] {rest of whole-cycle} (4);

\end{tikzpicture}

  \end{figure}
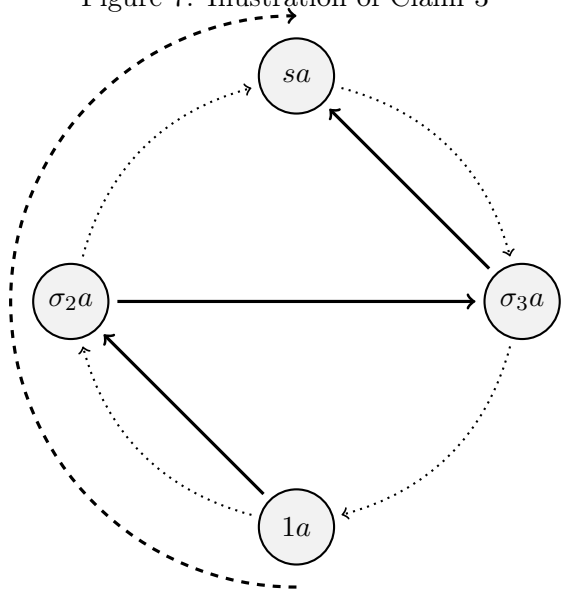
  \begin{claim}
    \label{claim:neglength}
    If the arc $(sa,1a)$ exists, it has length negative of the whole-cycle path from $1a$ to $sa$.
    That is, $\ell(sa,1a)=-\ell(1a,2a,...,\left(s-1\right)a,sa)$.
  \end{claim}
  From Claim \ref{claim:samelength}, $\ell(sa,1a)=\ell(sa,(s+1)a,...,Sa,1a)$. Notice that
  $(sa,(s+1)a,...,Sa,1a)$ and $(1a,2a,...,(s-1)a,sa)$ form the whole cycle, so their lengths sum to
  0. That is, $\ell(sa,1a)+\ell(1a,2a,...,(s-1)a,sa)=0$, and the claim follows.
  \begin{rem}
    The indexing of $1a$ and $sa$ in Claims \ref{claim:samelength} and \ref{claim:neglength} is not
    important. Since the whole-cycle is a cycle, $1a$ can be any vertex. (It is convenient to have
    $1\leq s\leq S$.)
  \end{rem}
  \begin{claim}
    \label{claim:anypath}
    Any $(1a,sa)$-path is the same length as the whole-cycle path
    $(1a,\underset{:=\alpha}{\underbrace{2a,...,\left(s-1\right)a}},sa)$.
  \end{claim}
  The $(1a,sa)$-path is some permutation of a subset of vertices of the SCC. Denote this
  $(\underset{=1a}{\underbrace{\sigma_{1}a}},\sigma_{2}a,...,\sigma_{j-1}a,\underset{=sa}{\underbrace{\sigma_{j}a}})$,
  where $j\leq S$. I will show
  \[
    \ell(\sigma_{1}a,...,\sigma_{j-1}a,\sigma_{j}a)=\underset{\text{whole-cycle path}}{\underbrace{\ell(1a,2a)+\cdots+\ell((\sigma_{j}-1)a,\sigma_{j}a)}}\equiv\sum_{i=1}^{\sigma_{j}-1}\ell(ia,(i+1)a)
  \]
  Note that $\sigma_{j-1}\not=\sigma_{j}-1$ in general.

  I will show the claim by strong induction on the length of $j$. The base case of $j=1$ is Claim
  \ref{claim:samelength}. Now suppose the claim is true for $j$; that is,
  $\ell(1a,...,\sigma_{j-1}a,\sigma_{j}a)=\sum_{i=1}^{\sigma_{j-1}}\ell(ia,(i+1)a)$. Now consider
  $j+1$. We have $\ell(1a,\sigma_{j+1}a)=\ell(1a,\sigma_{j}a)+\ell(\sigma_{j}a,\sigma_{j+1}a)$. If
  $\sigma_{j+1}>\sigma_{j}$, then by Claim \ref{claim:samelength} write
  \begin{align*}
    \ell(\sigma_{j}a,\sigma_{j+1}a) & =\sum_{i=\sigma_{j}}^{\sigma_{j+1}-1}\ell(ia,(i+1)a)
  \end{align*}
  So
  \begin{align*}
    \ell(1a,...,\sigma_{j}a,\sigma_{j+1}a) & =\sum_{i=1}^{\sigma_{j}-1}\ell(ia,(i+1)a)+\ell(\sigma_{j}a,\sigma_{j+1}a)                     \\
                                           & =\sum_{i=1}^{\sigma_{j}-1}\ell(ia,(i+1)a)+\sum_{i=\sigma_{j}}^{\sigma_{j+1}-1}\ell(ia,(i+1)a) \\
                                           & =\sum_{i=1}^{\sigma_{j+1}-1}\ell(ia,(i+1)a)
  \end{align*}
  If $\sigma_{j+1}<\sigma_{j}$, then by Claim \ref{claim:neglength} write
  \begin{align*}
    \ell(\sigma_{j}a,\sigma_{j+1}a) & =-\sum_{i=\sigma_{j}}^{\sigma_{j+1}-1}\ell(ia,(i+1)a)
  \end{align*}
  So
  \begin{align*}
    \ell(1a,...,\sigma_{j}a,\sigma_{j+1}a) & =\sum_{i=1}^{\sigma_{j}-1}\ell(ia,(i+1)a)+\ell(\sigma_{j}a,\sigma_{j+1}a)                                                                \\
                                           & =\sum_{i=1}^{\sigma_{j+1}-1}\ell(ia,(i+1)a)+\sum_{i=1}^{\sigma_{j}-1}\ell(ia,(i+1)a)-\sum_{i=\sigma_{j}}^{\sigma_{j+1}-1}\ell(ia,(i+1)a) \\
                                           & =\sum_{i=1}^{\sigma_{j+1}-1}\ell(ia,(i+1)a)
  \end{align*}
  as desired.

  \begin{figure}[H]
    \centering
    \caption{Illustration of Claim \ref{claim:anypath}}
    \label{fig:anypath}



\begin{tikzpicture}[
roundnode/.style={circle, draw=black, fill=black!5, thick, minimum size=10mm},
emptynode/.style={circle, draw=white, fill=white, thick, minimum size=5mm},
scale=1
]

\node[roundnode]    (1)    at (-3,0)    {$\sigma_2 a$};
\node[roundnode]    (2)    at (3,0)    {$\sigma_3 a$};

\node[roundnode]    (end)    at (0,3)    {$sa$};
\node[roundnode]    (start)    at (0,-3)    {$1a$};

\path [thick, ->, shorten >=3pt, shorten <=3pt]
    (start) edge [very thick, ]	(1) 
    (1) edge [very thick, ]		(2) 
    (2) edge [very thick, ]		(end) 
    
    (start) edge [bend left, dotted]	(1)
    (1) edge [bend left, dotted]	(end)
    (end) edge [bend left, dotted]	(2)
    (2) edge [bend left, dotted]	(start);
        
\draw [dashed, very thick, ->] (0,-3.8) arc(270:90: 3.8) ;

\end{tikzpicture}

  \end{figure}
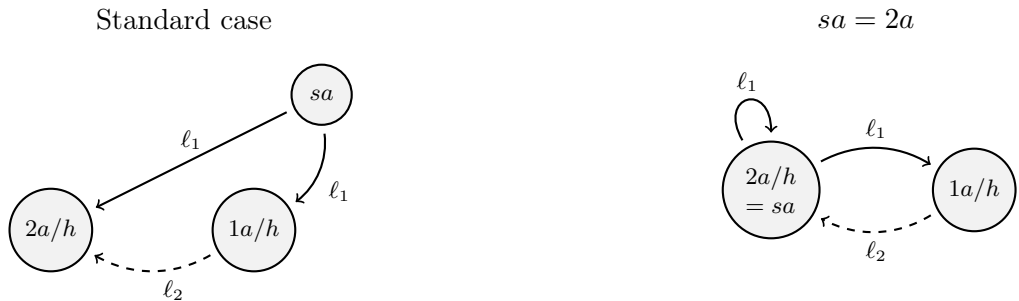
  \begin{claim}
    \label{claim:hlen0}
    The length of any path between an object type $h$ and itself is 0.
  \end{claim}
  Figure \ref{fig:hlen0} illustrates the following argument. Note that two vertices (agents) may be
  endowed with the same object type, so these can be distinct nodes. Recall that all copies of the
  same object type are contained in the same SCC. The path length from a vertex to itself is 0 since
  the whole-cycle has length 0, and any other path is the same length. Now suppose $h$ is contained
  in two distinct vertices, $1a$ and $2a$. Consider a node $sa$ such that $x_{sa}=h$. (This may be
  $1a$ or $2a$.) Then the arcs $(sa,1a)$ and $(sa,2a)$ exist. These have the same length,
  $\omega_{sa}-m_{sa}$, by construction of $\mathcal{G}_{T}$. Denote
  $\ell(sa,1a)=\ell(sa,2a)=\ell_{1}$. I show the length of the path from $1a$ to $2a$ is 0. Denote
  this path $(1a,...,2a)$, and let $\ell(1a,...,2a)=\ell_{2}$. Both $(sa,1a,...,2a)$ and $(sa,2a)$
  are paths from $sa$ to $2a$, so must have the same length. Then $\ell_{1}=\ell_{1}+\ell_{2}$,
  giving us $\ell_{2}=0$ as desired.

  \begin{figure}[H]
    \centering
    \caption{Illustration of Claim \ref{claim:hlen0}}
    \label{fig:hlen0}
    \begin{minipage}[t]{0.4\columnwidth}%
      \begin{center}
        Standard case
      \end{center}
      \begin{center}


\begin{tikzpicture}[
roundnode/.style={circle, draw=black, fill=black!5, thick, minimum size=8mm},
scale=0.9
]

\tikzstyle{every node}=[font=\footnotesize]

\node[roundnode]    (1)    at (4,2)    {$sa$};
\node[roundnode]    (2)    at (3,0)    {$1a/h$};
\node[roundnode]    (3)    at (0,0)    {$2a/h$};

\path [thick, ->, shorten >=3pt, shorten <=3pt]
    (1) edge [bend left] node[below right] {$\ell_1$} (2) 
    (2) edge [bend left, dashed] node[below right] {$\ell_2$} (3) 
    (1) edge [] node[above] {$\ell_1$}  (3) ;

\end{tikzpicture}

      \end{center}%
    \end{minipage}\hfill%
    \begin{minipage}[t]{0.4\columnwidth}%
      \begin{center}
        $sa=2a$
      \end{center}
      \begin{center}


\begin{tikzpicture}[
roundnode/.style={circle, draw=black, fill=black!5, thick, minimum size=8mm},
scale=0.9,
align=left
]

\tikzstyle{every node}=[font=\footnotesize]

\node[roundnode]    (2)    at (3,0)    {$1a/h$};
\node[roundnode]    (3)    at (0,0)    {$2a/h$ \\ $=sa$} ;

\path [thick, ->, shorten >=3pt, shorten <=3pt]
    (3) edge [bend left] node[above] {$\ell_1$} (2) 
    (2) edge [bend left, dashed] node[below] {$\ell_2$}  (3) 
    (3) edge [loop, looseness=6, out=120, in=90] node[above]{ $\ell_1$} (3); 

\end{tikzpicture}

      \end{center}%
    \end{minipage}
  \end{figure}

  I have shown the above claims for SCCs of size at least three. Now consider an SCC of only one
  vertex. The only arc must be $(1a,1a)$, which constitutes the whole-cycle and must have length 0,
  and the path length from this object type to itself is 0.

  Now consider an SCC of two vertices, $1a$ and $2a$. If they are endowed with distinct object
  types, the arcs $(1a,2a)$ and $(2a,1a)$ are the only arcs, and the claims are true trivially. If
  they are endowed with the same object type, the self loops are also present. The two self-loops
  partition the SCC, so have length 0. We have $\ell(1a,1a)=\ell(1a,2a)$ by construction, so
  $\ell(1a,2a)=0$, and similarly $\ell(2a,1a)=0$. Then all arcs have length 0 in this SCC, so the
  claims are again true.

  The rest of the proof follows easily. The path length between any object type $h$ and itself is 0
  (so the minimum path length is 0), ensuring it is possible to construct prices this way. Next, for
  any $ik\in\mathcal{A}$, the path length from $e_{ik}:=h$ to $x_{ik}:=h'$ is $m_{ik}-\omega_{ik}$,
  so that $p_{h'}-p_{h}=m_{ik}-\omega_{ik}$. This gives
  \[
    (x_{ik}-e_{ik})\cdot p=p_{h'}-p_{h}=m_{ik}-\omega_{ik}
  \]
  as desired.

  This completes the proof of the theorem.
\end{proof}
\begin{proof}[Proof of Corollary \ref{cor:strictTUrat}.]
  As argued in the proof of Theorem \ref{thm:MainTU}, any price must satisfy
  $(x_{ik}-e_{ik})\cdot p=\omega_{ik}-m_{ik}$ for all $ik\in\mathcal{A}$. By the construction of
  $\mathcal{G}_{T}$, $x_{ik}-e_{ik}$ is an arc from $e_{ik}$ to $x_{ik}$ with length
  $\omega_{ik}-m_{ik}$, which is also the price difference between these objects. Inductively (I
  will omit the full formality), a path from $x_{ik}$ to $x_{ik'}$ has path length 0 if and only if
  the price difference between them is 0. (Note that by Claim \ref{claim:neglength}, there also must
  be a path from $x_{ik'}$ to $x_{ik}$, and it has length 0 as well.)

  (``If'') Let both conditions be true. As in the main theorem, it is sufficient to set
  $v_{i}(x_{ik})=p\cdot x_{ik}$. Since prices can be set arbitrarily across SCCs, we can ensure no
  two objects in different SCCs have the same price.

  (``Only if'') Toward a contradiction, suppose the economy is not T-rationalizable. Then it is of
  course not strictly T-rationalizable. Now suppose the second condition is false. That is, there
  are $ik,ik'$ in the same SCC such that $x_{ik}\not=x_{ik'}$, but the shortest path length between
  them is 0. Then $p_{x_{ik}}=p_{x_{ik'}}$. Suppose $v_{i}(x_{ik})>v_{i}(x_{ik'})$ without loss of
  generality. Then $ik'$ can afford $(x_{ik},m_{ik'})$, which is preferable to $(x_{ik'},m_{ik'})$.
  Thus $(x,m)$ is not a competitive equilibrium, so is not strictly T-rationalizable.

  In particular, $x_{ik}$ can purchase $x_{ik'}$ instead. Since $m_{ik}>0$ by assumption, $ik$ can
  form a blocking coalition by compensating other members of the blocking coalition.
\end{proof}
\begin{proof}[Proof of Corollary \ref{cor:Afriat}.]
  This comes from the proof of Theorem \ref{thm:MainTU}.

  The first line is conditions for valid vectors $p$, which comes from Theorem \ref{thm:MainTU} and
  its proof.

  The first inequality is $(A)$. This is exactly Afriat's inequalities when the marginal utility of
  money is 1. These give joint restrictions on any the utility for objects actually consumed by
  agent type $i$ given some $p$. Necessity and sufficiency are from Afriat's theorem.

  The second inequality gives restrictions on the utility for objects not consumed by type $i$. An
  object $h$ that is affordable under some $ik$'s budget must have
  $V(h,e_{ik}\cdot p+\omega_{ik}-p\cdot h)\leq V(x_{ik},e_{ik}\cdot p+\omega_{ik}-p\cdot x_{ik})$,
  else $(x,m)$ is not a competitive equilibrium. This gives the inequality in the corollary:
  \begin{align*}
    v_{i}(h)+(e_{ik}\cdot p+\omega_{ik}-h\cdot p) & \leq v_{i}(x_{ik})+(e_{ik}\cdot p+\omega_{ik}-x_{ik}\cdot p) \\
    v_{i}(h)-h\cdot p                             & \leq v_{i}(x_{ik})-x_{ik}\cdot p
  \end{align*}
  That is, if $h$ is affordable to $ik$, then its utility (including leftover money) must be less
  than that of $x_{ik}$. Note that an object that is too expensive for all $ik$ is allowed to have
  any utility. Again, necessity and sufficiency are immediate.
\end{proof}

\end{document}